\documentclass[usenatbib]{mn2e} 
\usepackage{graphicx,fixltx2e,amsmath,amssymb,amsfonts,latexsym,wasysym}

\def\chisq{\hbox{$\chi^2$}}
\def\chisqr{\hbox{$\chi^2_{\rm r}$}}
\def\msun{\hbox{${\rm M}_{\odot}$}}

\def\mspy{\hbox{${\rm M}_{\odot}$\,yr$^{-1}$}}
\def\rsun{\hbox{${\rm R}_{\odot}$}}
\def\lsun{\hbox{${\rm L}_{\odot}$}}
\def\rcor{\hbox{$r_{\rm cor}$}}
\def\rmag{\hbox{$r_{\rm mag}$}}
\def\mstar{\hbox{$M_{\star}$}}
\def\rstar{\hbox{$R_{\star}$}}

\def\teff{\hbox{$T_{\rm eff}$}}
\def\logg{\hbox{$\log g$}}
\def\sn{\hbox{S/N}}

\def\kms{\hbox{km\,s$^{-1}$}}
\def\vsini{\hbox{$v \sin i$}}

\def\mV{\hbox{$m_{\rm V}$}}
\def\AV{\hbox{$A_{\rm V}$}}

\def\degr{\hbox{$^\circ$}}

\newcommand{\caii}{\hbox{Ca$\;${\sc ii}}}
\newcommand{\fei}{\hbox{Fe$\;${\sc i}}}

\newcommand{\hei}{\hbox{He$\;${\sc i}}}
\newcommand{\hal}{\hbox{H${\alpha}$}}
\newcommand{\hbe}{\hbox{H${\beta}$}}

\begin{document}

\title[Magnetospheric accretion on the cTTS DN~Tau]{Magnetospheric accretion on the fully-convective classical T~Tauri star DN~Tau} 
\makeatletter

\def\newauthor{%
  \end{author@tabular}\par
  \begin{author@tabular}[t]{@{}l@{}}}
\makeatother
 
\author[J.-F.~Donati et al.]
{\vspace{1.7mm}
J.-F.~Donati$^1$\thanks{E-mail: 
jean-francois.donati@irap.omp.eu }, 
S.G.~Gregory$^2$, S.H.P.~Alencar$^3$, G.~Hussain$^4$, J.~Bouvier$^5$, \\  
\vspace{1.7mm}
{\hspace{-1.5mm}\LARGE\rm 
M.M.~Jardine$^2$, F.~M\'enard$^{6,5}$, C.~Dougados$^{6,5}$, M.M.~Romanova$^7$ \& the MaPP}\\ 
\vspace{1.7mm}
{\hspace{-1.5mm}\LARGE\rm
collaboration} \\
$^1$ UPS-Toulouse / CNRS-INSU, Institut de Recherche en Astrophysique et Plan\'etologie (IRAP) UMR 5277, Toulouse, F--31400 France \\ 
$^2$ School of Physics and Astronomy, Univ.\ of St~Andrews, St~Andrews, Scotland KY16 9SS, UK \\ 
$^3$ Departamento de F\`{\i}sica -- ICEx -- UFMG, Av. Ant\^onio Carlos, 6627, 30270-901 Belo Horizonte, MG, Brazil \\ 
$^4$ ESO, Karl-Schwarzschild-Str.\ 2, D-85748 Garching, Germany \\ 
$^5$ UJF-Grenoble 1 / CNRS-INSU, Institut de Planétologie et d'Astrophysique de Grenoble (IPAG) UMR 5274, Grenoble, F-38041, France \\ 
$^6$ UMI-FCA, CNRS/INSU, France (UMI 3386), and Universidad de Chile, Santiago, Chile \\ 
$^7$ Department of Astronomy, Cornell University, Ithaca, NY 14853-6801, USA  
}

\date{2013 August, MNRAS accepted}
\maketitle
 
\begin{abstract}  
We report here results of spectropolarimetric observations of the classical T~Tauri star DN~Tau
carried out (at 2 epochs) with ESPaDOnS at the Canada-France-Hawaii Telescope within the
`Magnetic Protostars and Planets' programme.
We infer that DN~Tau, with a photospheric temperature of $3,950\pm50$~K, a luminosity of
$0.8\pm0.2$~\lsun\ and a rotation period of 6.32~d, is a $\simeq$2~Myr-old fully-convective
$0.65\pm0.05$~\msun\ star with a radius of $1.9\pm0.2$~\rsun, viewed at an inclination of $35\pm10$\degr.

Clear circularly-polarized Zeeman signatures are detected in both photospheric and accretion-powered
emission lines, probing longitudinal fields of up to 1.8~kG (in the \hei\ $D_3$ accretion proxy).
Rotational modulation of Zeeman signatures, detected both in photospheric and accretion lines,
is different between our 2 runs, providing further evidence that fields of cTTSs are generated
by non-stationary dynamos.

Using tomographic imaging, we reconstruct maps of the large-scale field, of the photospheric
brightness and of the accretion-powered emission at the surface of DN~Tau at both epochs.
We find that the magnetic topology is mostly poloidal, and largely axisymmetric, with
an octupolar component (of polar strength 0.6--0.8~kG) 1.5--2.0$\times$ larger than the 
dipolar component (of polar strength $\simeq$0.3--0.5~kG).  

DN~Tau features dominantly poleward accretion at both epochs.  The large-scale
dipole component of DN~Tau is however too weak to disrupt the surrounding accretion disc
further than 65--90\% of the corotation radius (at which the disc Keplerian period matches
the stellar rotation period), suggesting that DN~Tau is already spinning up despite being fully convective.
\end{abstract}


\section{Introduction} 
\label{sec:int}

Magnetic fields have a significant impact on the life of low-mass stars.  For instance, they are known to 
give rise to various activity phenomena throughout the atmospheres of cool stars (depending on their mass and 
rotation rates), including dark spots and bright plages at stellar surfaces, high-temperature outer atmospheric 
layers (e.g., chromospheres and coronae), energetic flares resulting from reconnection events in the 
magnetosphere, and mass ejections and winds escaping the stars (both a direct consequence of their coronae).  
Through these activity processes, magnetic fields can in particular drastically brake 
the rotation rates of Sun-like stars in a few 0.1~Gyrs, with large-scale fields expelling angular momentum outwards 
by forcing the inner wind regions to co-rotate with stellar surfaces;  for this reason, most single cool dwarfs 
rapidly end up as weakly-active, slowly-rotating Sun-like stars and remain such for most of their time on the main sequence.  

During formation stages, the role of magnetic fields on the evolution of low-mass stars is far larger than it is 
for mature cool dwarfs, getting comparable in scale to that played by turbulence and immediately behind the prime 
role played by gravitation \citep[e.g.,][]{Andre09}.  
For instance, interstellar magnetic fields can slow down the initial collapse of parsec-sized molecular clouds 
as they are turned into pre-stellar cores;  they can also strongly affect the formation, the radial structure and 
the accretion rate of the accretion discs surrounding the newly-born protostars, and are held responsible 
for producing disc outflows of various types (e.g., collimated jets, conical winds, magnetic towers) through 
which accretion discs evacuate a significant fraction of the initial cloud material and angular momentum.  

At a later formation stage, low-mass protostars that still possess a massive accretion disc (the so-called classical 
T~Tauri stars or cTTSs) host magnetic fields strong enough to disrupt the central regions of their discs 
and to control accretion through discrete funnels linking the inner discs to the stellar surfaces. 
In this process, magnetic fields can brake the rotation of cTTSs via a star/disc coupling mechanism whose physical 
nature is still a matter of speculation \citep[e.g.,][]{Bouvier07, Mohanty08, Romanova11, Matt12, Zanni13}.  
Photometric monitorings of stellar formation regions \citep[e.g.,][]{Irwin09} indeed demonstrate that the most 
slowly-rotating members of these regions, presumably the cTTSs, manage to maintain their slow rotation rates in the 
first few Myr of their lives, and despite both sustained contraction and accretion, at least when their masses 
are larger than 0.5~\msun\ typically.  Very-low-mass protostars are found to behave rather differently;  
their rotation rates steadily increase over the same period of time, suggesting that the braking scheme implemented 
by their higher-mass analogs is much less efficient at very-low masses.  Detecting and measuring magnetic 
fields of cTTSs, and looking at how they vary with masses, ages and rotation rates, should thus 
bring a wealth of novel physical information regarding the angular momentum evolution of newly-born Sun-like stars.  

However, magnetic fields of protostars are tricky to estimate, especially their large-scale topologies that are 
expected to play the main role in star-disc coupling mechanisms.  All such measurements are presently obtained 
by using the Zeeman effect on spectral lines, capable not only of distorting the profiles (and in particular the 
widths) of unpolarized spectral lines, but also of generating genuine circular and linear polarization signatures 
in line profiles, depending on the Zeeman sensitivities of the considered spectral lines and on the orientation of 
the magnetic field with respect to the line of sight \citep[e.g.,][]{Donati09}.  The first estimates, derived from 
the differential broadening of unpolarized spectral lines (especially in the near infrared), yielded magnetic strengths 
of several kG for a number of cTTSs \citep[e.g.,][]{Johns99b, Johns07};  however, being poorly sensitive to magnetic 
topologies, these initial studies can hardly help regarding the role of magnetic fields in slowing down the rotation 
of cTTSs.  By focussing on sets of circularly-polarized Zeeman signatures of cTTSs, recent studies managed to retrieve 
the large-scale magnetic topologies at the surfaces of low-mass protostars thanks to tomographic imaging 
techniques \citep{Donati07, Donati08}, thus opening a new option for investigating the issue in a more direct, observational way.  

MaPP (Magnetic Protostars and Planets) is a project aiming at measuring the large-scale topologies in a small sample 
of about 15 prototypical cTTSs of various masses and ages, and at assessing, on the basis of theoretical modelling, 
whether and how these fields are able to control / contribute to the observed dissipation of angular momentum of 
cTTSs.  A total of 640~hr of telescope time was allocated to MaPP on the 3.6~m Canada-France-Hawaii Telescope (CFHT) 
with the ESPaDOnS high-resolution spectropolarimeter, over a timescale of 9 semesters (2008b-2012b) - the collected 
data set being now complete.  Regarding low- and intermediate-mass cTTSs (with masses ranging from 0.35 to 1.35~\msun), 
8 stars have been studied in detail so far (and in some cases at various epochs) using an upgraded version of the 
imaging code \citep{Donati10b}.  The current picture emerging from these observations is that large-scale topologies of 
cTTSs, being generated by (non-stationary) dynamo processes, largely reflect the stellar internal structures and in 
particular the degree to which the stars are convective \citep{Donati12, Gregory12};  moreover, the strong dipole 
components in the large-scale fields of a few cTTSs may provide a natural explanation for the slow rotation of 
protostars young enough to be fully-convective \citep[e.g.,][]{Zanni13}.  

With only one very-low mass cTTS studied yet \citep[V2247~Oph,][whose magnetic properties are dissimilar to those of 
other fully-convective stars]{Donati10}, it is not clear yet how these first conclusions apply to stars with masses 
lower than 0.7~\msun\ and what it means in terms of their apparent inability to counteract their natural spin up.  
We therefore focused, for this new study, on DN~Tau, one of the 2 lowest-mass stars of our cTTS sample.  
After a brief description of our dual-epoch MaPP observations (Sec.~\ref{sec:obs}) and a short summary of what is 
known (and relevant to this paper) on DN~Tau (Sec.~\ref{sec:dn}), we describe in more details the rotational 
modulation and intrinsic variability we observe (Sec.~\ref{sec:var}), carry-out the modelling of the large-scale 
magnetic field, brightness and accretion maps of DN~Tau (Sec.~\ref{sec:mod}) and summarise the results and their 
implications for our understanding of magnetospheric accretion processes of cTTSs and their impact on the angular 
momentum evolution of low-mass stars (Sec.~\ref{sec:dis}).

\section{Observations}
\label{sec:obs}

Spectropolarimetric observations of DN~Tau were collected at two different epochs, using the high-resolution 
spectropolarimeters ESPaDOnS at the 3.6-m Canada-France-Hawaii Telescope (CFHT) atop Mauna Kea (Hawaii) 
and its brothership NARVAL at the 2-m T\'elescope Bernard Lyot (TBL) atop Pic du Midi (France).  
Both ESPaDOnS and NARVAL collect stellar spectra spanning the entire optical domain (from 370 to 1,000~nm) 
at a resolving power of 65,000 (i.e., resolved velocity element of 4.6~\kms), in either circular or linear 
polarisation \citep{Donati03} - only circular polarisation being used in the present study. 
The first data set, referred to as 2010~Dec in the following, was collected from 2010 Nov~27 
to 2011~Jan~03 using both ESPaDOnS and NARVAL;  the second data set, called 2012~Dec herein, was collected 
from 2012 Nov~24 to Dec~10 using ESPaDOnS only (dreadful weather at TBL prevented the collection of  
useful data on DN~Tau).  

In 2010~Dec, a total of 13 circular polarisation spectra were collected over a timespan of 38 nights, 
corresponding to about 6 rotation cycles of DN~Tau;  the time sampling is irregular at the beginning and at the 
end of the run, reflecting the variable weather conditions on both sites (with, e.g., only 1 spectrum collected 
over the first 12 nights or $\simeq$2 first rotation cycles), but much more even (with 1 spectrum per night) in 
the middle of the run (around the third rotation cycle).  
In 2012~Dec, a total of 11 circular polarisation spectra were collected over 16 nights (2.5 rotation cycles), 
with a few gaps of 2 to 3 nights in the overall phase coverage.  

All polarisation spectra consist of 4 individual subexposures (each lasting about 1200~s) taken in different 
polarimeter configurations to allow the removal of all spurious polarisation signatures at first order.
All raw frames are processed as described in the previous papers of the series 
\citep[e.g.,][]{Donati10b, Donati11}, to which the reader is referred for more information.  
The peak signal-to-noise ratios (\sn, per 2.6~\kms\ velocity bin) achieved on the
collected spectra range between 160 and 230 with ESPaDOnS (with a median of 200) and between 100 and 120 with NARVAL, 
depending mostly on weather/seeing conditions.  
The full journal of observations is presented in Table~\ref{tab:log}.

\begin{table}
\caption[]{Journal of observations of DN~Tau collected in 2010--2011, and in 2012.
Each observation consists of a sequence of 4 subexposures, lasting either 1292~s (on 2010 Dec~14--16), 1250~s 
(on 2010 Dec~17--19 and Dec~26--30), 1238.5~s (in 2012 Dec) or 1200~s (with NARVAL).  
Columns $1-4$ respectively list (i) the UT date of the observation and the instrument used to collect it 
(N \& E for NARVAL and ESPaDOnS respectively), (ii) the Heliocentric Julian Date (HJD) 
in excess of 2,450,000 (at mid-exposure), (iii) the corresponding Heliocentric UT time, 
and (iv) the peak signal to noise ratio (per 2.6~\kms\ velocity bin) of each observation.
Column 5 lists the rms noise level (relative to the unpolarized continuum level
$I_{\rm c}$ and per 1.8~\kms\ velocity bin) in the circular polarization profile
produced by Least-Squares Deconvolution (LSD), while column~6 indicates the
orbital/rotational cycle associated with each exposure (using the ephemeris given by
Eq.~\ref{eq:eph}).  }
\begin{tabular}{cccccc}
\hline
Date & HJD          & UT      &  \sn\  & $\sigma_{\rm LSD}$ & Cycle \\
(2010/11) & (5,500+) & (h:m:s) &      &   (0.01\%)  & (0+) \\
\hline
Nov 27 (N) & 27.56612 & 01:27:42 & 100 & 5.2 & 0.090 \\
Dec 09 (N) & 40.52290 & 24:25:53 & 100 & 5.2 & 2.140 \\
Dec 10 (N) & 41.52669 & 24:31:23 & 100 & 5.3 & 2.299 \\
Dec 14 (E) & 44.91493 & 09:50:38 & 160 & 3.1 & 2.835 \\
Dec 15 (E) & 45.98117 & 11:26:04 & 200 & 2.3 & 3.003 \\
Dec 16 (E) & 46.98269 & 11:28:20 & 190 & 2.6 & 3.162 \\
Dec 17 (E) & 47.95388 & 10:46:54 & 190 & 2.5 & 3.315 \\
Dec 18 (E) & 48.94556 & 10:34:59 & 200 & 2.6 & 3.472 \\
Dec 19 (E) & 49.91588 & 09:52:18 & 170 & 3.0 & 3.626 \\
Dec 20 (N) & 51.44662 & 22:36:41 & 120 & 4.0 & 3.868 \\
Dec 26 (E) & 56.97137 & 11:12:44 & 210 & 2.4 & 4.742 \\
Dec 30 (E) & 60.91432 & 09:50:55 & 160 & 3.6 & 5.366 \\
Jan 03 (N) & 65.32907 & 19:48:33 & 110 & 4.5 & 6.065 \\
\hline
Date & HJD          & UT      &  \sn\  & $\sigma_{\rm LSD}$ & Cycle \\
(2012)  & (6,200+) & (h:m:s) &      &   (0.01\%)  & (115+) \\
\hline
Nov 24 (E) & 56.05719 & 13:14:49 & 200 & 2.4 & 0.357 \\
Nov 25 (E) & 57.04158 & 12:52:21 & 210 & 2.2 & 0.513 \\
Nov 27 (E) & 59.00792 & 12:03:54 & 210 & 2.2 & 0.824 \\
Nov 28 (E) & 60.01676 & 12:16:40 & 230 & 1.9 & 0.984 \\
Nov 29 (E) & 60.87053 & 08:46:07 & 190 & 2.3 & 1.119 \\
Dec 01 (E) & 63.07944 & 13:46:59 & 200 & 2.2 & 1.468 \\
Dec 04 (E) & 66.08836 & 13:59:57 & 230 & 1.9 & 1.944 \\
Dec 07 (E) & 68.96865 & 11:07:40 & 200 & 2.3 & 2.400 \\
Dec 08 (E) & 70.03873 & 12:48:38 & 210 & 2.1 & 2.569 \\
Dec 09 (E) & 71.01165 & 12:09:41 & 200 & 2.4 & 2.723 \\
Dec 10 (E) & 71.94625 & 10:35:33 & 200 & 2.2 & 2.871 \\
\hline
\end{tabular}
\label{tab:log}
\end{table}

Rotational cycles $E$ of DN~Tau are computed from Heliocentric Julian Dates (HJDs)
according to the following ephemeris:
\begin{equation}
\mbox{HJD} = 2455527.0 + 6.32 E   
\label{eq:eph}
\end{equation}
in which the rotation period is taken from the most recent literature \citep{Artemenko12} 
(in good agreement with our own determination, see Sec.~\ref{sec:var}) 
and the initial Julian date is chosen arbitrarily.  
As already mentioned, our data are collected over several successive rotation cycles of DN~Tau, 
offering us a convenient way of disentangling intrinsic variability from rotational modulation in
the spectra (see Sec.~\ref{sec:var})\footnote{The accuracy on the period is however not good enough 
to allow relating the phases of our 2010 spectra with those of the 2012 ones.}.    

\begin{figure}
\includegraphics[scale=0.35,angle=-90]{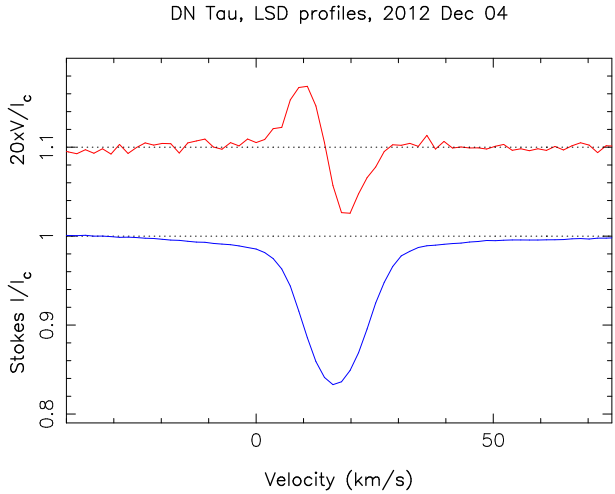}
\caption[]{LSD circularly-polarized (Stokes $V$) and unpolarized (Stokes $I$)
profiles of DN~Tau (top/red, bottom/blue curves respectively) collected on
2012~Dec 04 (cycle 115+1.944).  A clear Zeeman signature (with a full amplitude of 0.7\%)  
is detected in the LSD Stokes $V$ profile, in conjunction with the unpolarised line profile.
The mean polarization profile is expanded by a factor of 20 and shifted upwards
by 1.1 for display purposes.  }
\label{fig:lsd}
\end{figure}

Least-Squares Deconvolution \citep[LSD,][]{Donati97b} was applied to all
observations.   The line list we employed for LSD is computed from an {\sc
Atlas9} LTE model atmosphere \citep{Kurucz93} featuring $\teff=4,000$~K and $\logg=3.5$, 
appropriate for DN~Tau (see Sec.~\ref{sec:dn}).
As usual, only moderate to strong atomic spectral lines are included
in this list \citep[see, e.g.,][for more details]{Donati10b}.  

Altogether, about 9,200 spectral features (with about 35\% from \fei) are used
in this process.
Expressed in units of the unpolarized continuum level $I_{\rm c}$, the average
noise levels of the resulting Stokes $V$ LSD signatures range from 1.9 to 3.6$\times10^{-4}$ 
per 1.8~\kms\ velocity bin (median value 2.3$\times10^{-4}$) with ESPaDOnS spectra, and 
about twice as much with NARVAL spectra.  
Zeeman signatures are detected at all times in LSD profiles and in our two main accretion
proxies (see Sec.~\ref{sec:var});  an example LSD photospheric Zeeman signature
(collected during the 2011 run) is shown in Fig.~\ref{fig:lsd} as an
illustration.

\section{Evolutionary stage of DN~Tau}
\label{sec:dn}

The photospheric temperature of DN~Tau \citep[spectral type M0,][]{Cohen79} is often quoted as 3850~K 
in the literature \citep[e.g.,][]{Kenyon95}, on account of the reported spectral type.  Direct estimates 
derived from multicolour photometry, lead to similar values \citep[e.g., 3920~K, e.g.,][]{Bouvier89}.  
Although not mentioned in the corresponding papers, the uncertainties on these estimates are likely to be at least 100~K.  

For the present study, we derived a new estimate directly from our spectra of DN~Tau thanks to the automatic 
spectral classification tool especially developed in the context of MaPP, inspired from that of \citet{Valenti05}, 
and already used and discussed in a previous paper of the series \citep{Donati12}.  Not only is this new estimate 
expected to be more accurate than the older ones, but it should also have the advantage of being homogeneous with 
measurements obtained with the same tool on all other cTTSs of our sample.  Applying it to our spectra of DN~Tau, 
we obtain $\teff=3,950\pm50$~K and $\logg=3.7\pm0.1$, slightly higher though still compatible with older 
literature estimates.  We use this value throughout our paper.  

To work out the luminosity of DN~Tau, we start from its brightest $V$ magnitude as observed during long-term 
photometric monitoring, equal to $\mV=12.1$ \citep[e.g.,][]{Artemenko12}.  We first note that the visual extinction 
that DN~Tau suffers is only moderate \citep[$\AV\simeq0.5$,][]{Kenyon95} and more or less compensates the limited 
continuum contribution from accretion veiling (see Sec.~\ref{sec:var}), at least to a precision of about 0.25~mag - 
qualitatively consistent with the approximate agreement found between spectroscopic and photometric temperature estimates.  
Given the temperature of DN~Tau, the required bolometric correction to apply is $-1.10$ \citep{Bessell98};  for a 
distance to Taurus of 140~pc, the correction to apply (distance modulus) is equal to $-5.73$.  Taking finally into 
account that spots (either always in view, or evenly spread over the surface) are very often present on active stars 
(and thus on cTTSs as well), typically reducing their luminosity by 20\% (even when brightest), we end up for DN~Tau 
with a bolometric magnitude of $5.0\pm0.25$, i.e., with a logarithmic luminosity (with respect to the Sun) of $-0.1\pm0.1$.  
The revised mass that we derive for DN~Tau \citep[mostly from \teff\ and a comparison with evolutionary models of][see 
Fig.~\ref{fig:hrd}]{Siess00} is $\mstar=0.65\pm0.05$~\msun\ (the error bar reflecting mainly the uncertainty on \teff);  
the corresponding radius is $\rstar=1.9\pm0.2$~\rsun, in good agreement with previous estimates 
\citep[e.g.,][]{Appenzeller05}.  The age we infer for DN~Tau is $\simeq$2~Myr, clearly indicating that DN~Tau 
is still fully convective (see Fig.~\ref{fig:hrd}).  

\begin{figure}
\includegraphics[scale=0.35,angle=-90]{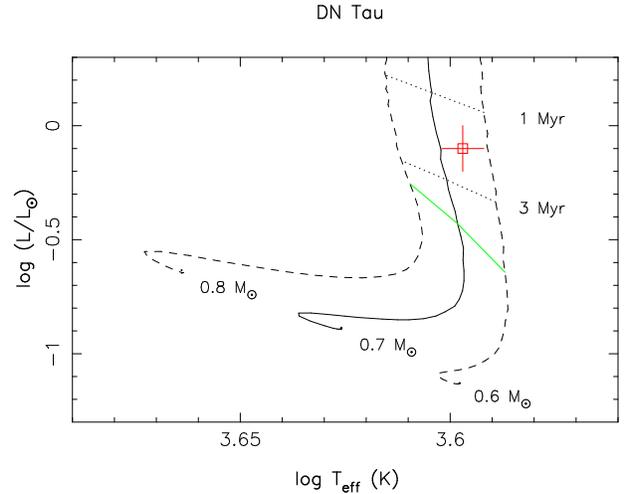}
\caption[]{Observed (open square and error bars) location of DN~Tau in the HR diagram.
The PMS evolutionary tracks and corresponding isochrones \citep{Siess00} assume solar
metallicity and include convective overshooting.  The green line depicts where models
predict cTTSs start developing their radiative core as they contract towards the main
sequence.  }
\label{fig:hrd}
\end{figure}

The rotation period of DN~Tau has been estimated a number of times, mostly from photometric monitoring \citep{Vrba86, 
Bouvier89, Percy10, Artemenko12};  the latest and most accurate of these studies yield a rotation period of 6.32~d 
(with which we phased all our data, see Sec.~\ref{sec:obs}), 
in relatively good agreement with previous estimates (ranging from 6.0 to 6.6~d).  Recent estimates obtained by 
monitoring the radial velocity (RV) of DN~Tau through cross-correlation of its photospheric spectrum 
\citep[e.g.,][]{Crockett12} confirm this periodicity and interpret the observed RV variations as rotational modulation 
of line profiles induced by the presence of large cool spots at the surface of DN~Tau, implying at the same time that 
the detected period truly relates to the rotation of the star (rather than to some variability in the inner regions of 
the accretion disc).  Our observations fully confirm this view (see Sec.~\ref{sec:var}).  

Given the line-of-sight-projected equatorial rotation velocity \vsini\ of DN~Tau that we derive \citep[equal to $9\pm1$~\kms, 
see Sec.~\ref{sec:mod}, in agreement with previous published estimates, e.g.,][]{Appenzeller05}, we obtain (from the radius 
and period estimated quoted above) that the rotation axis of DN~Tau is inclined at $35\pm10\degr$ to the line of sight, i.e., 
more or less coincident with that of the accretion disc \citep[e.g.,][]{Muzerolle03}.

\section{Spectroscopic variability}
\label{sec:var}

We now describe the spectroscopic variability that DN~Tau exhibited during our two observing runs, 
concentrating mostly on photospheric LSD profiles and on a small selection of accretion proxies 
(i.e., \caii\ infrared triplet / IRT, \hei\ $D_3$, \hal\ and \hbe).  In this step, we essentially focus on simple 
observables that can be derived from these profiles, and in particular on equivalent widths, RVs 
and longitudinal magnetic fields (i.e., the line-of-sight projected component of the vector field 
averaged over the visible stellar hemisphere and weighted by brightness inhomogeneities), and look 
at how they vary with time - both in terms of rotational modulation and intrinsic variability.  
This basic approach allows in particular to get an intuitive understanding of how the various 
observables support the existence of the multiple magnetic, brightness and accretion features 
present at the surface of DN~Tau.  

\subsection{LSD photospheric profiles}

\begin{figure*}
\center{\includegraphics[scale=0.85]{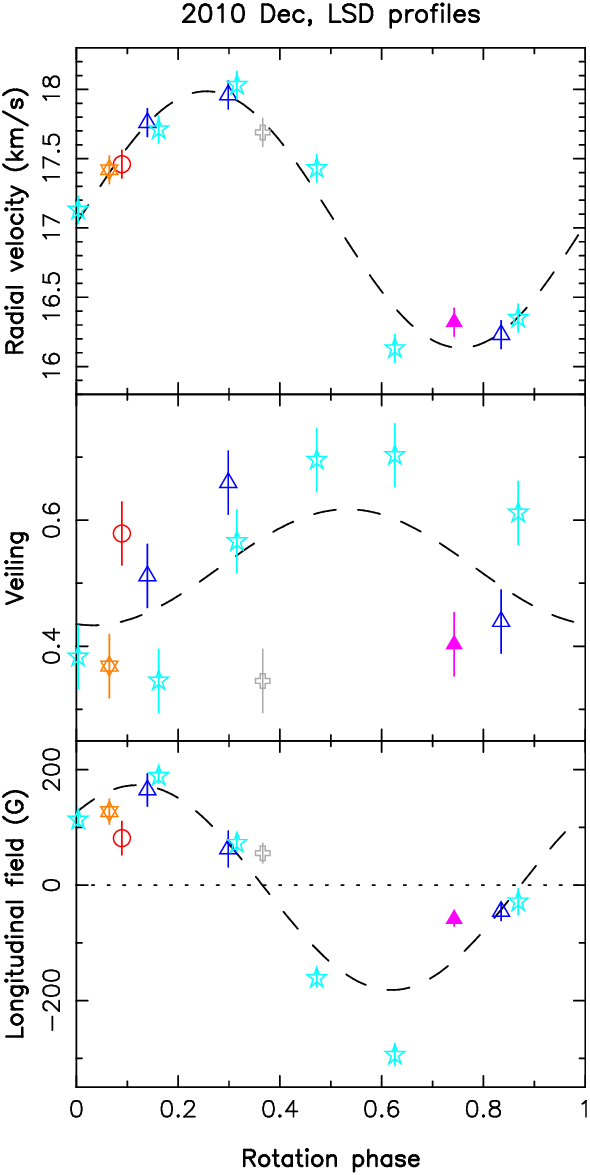}\hspace{4mm}
\includegraphics[scale=0.85]{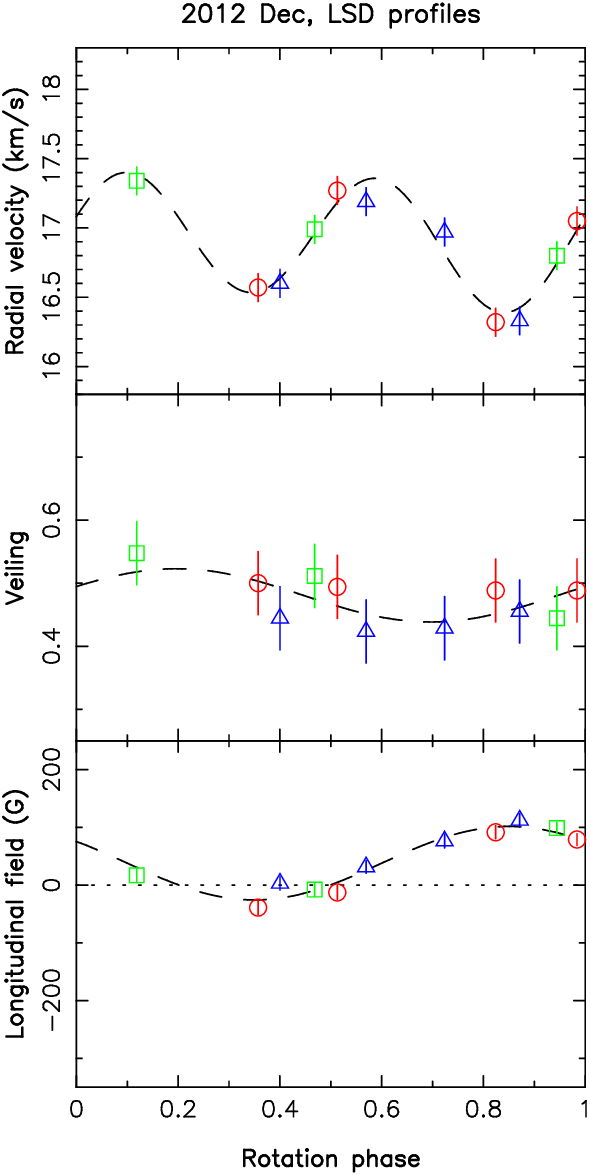}}
\caption[]{Rotational modulation of the RV (top row), veiling (second row)
and longitudinal field (bottom row) derived from the LSD photospheric profiles
of DN~Tau in late 2010 (left panels) and late 2012 (right panels).
Data collected during the successive rotation cycles of each run are shown with
different symbols and colors (red open circles, green open squares, dark-blue open triangles, 
light-blue open five-pointed stars, pink filled triangles, grey open plusses and orange 
open six-pointed stars respectively indicating rotation cycles 0 to 6).  
Formal $\pm$1~$\sigma$ error bars (computed from the error bars of the observed spectra) 
are shown for longitudinal fields, while conservative error bars of $\pm$0.1~\kms\ and 
$\pm$0.05 were assumed for RVs and veiling respectively.
Fits with sine/cosine waves (and first harmonic, whenever appropriate) are included 
(and shown as dashed lines) to outline (whenever significant) the amount of variability 
attributable to rotational modulation.  This figure is best viewed in color.  }
\label{fig:var1}
\end{figure*}

The temporal variability of unpolarized and circularly-polarized LSD profiles of DN~Tau is 
summarised in Fig.~\ref{fig:var1}, showing obvious differences between our two observing runs.  

In 2010~Dec (left column of Fig.~\ref{fig:var1}, top panel), RVs of unpolarized LSD profiles exhibit a 
conspicuous and simple rotational modulation about a mean RV of $\simeq17.1\pm0.1$~\kms\ and 
with a full amplitude of $\simeq$1.9~\kms;  this modulation repeats rather well across the 6 
rotation cycles, showing only a low level of intrinsic variability.  
As for most cTTSs studied to date in the MaPP sample, and in agreement with the findings 
of \citet{Crockett12}, this suggests the presence of a dark spot at the surface of DN~Tau, 
centred at phase 0.50 (i.e., at mid-phase between RV maximum and minimum) and located at high 
latitudes (given the fairly sinusoidal shape of the RV curve and its small semi-amplitude with 
respect to \vsini).  The period on which RVs fluctuate is $6.33\pm0.05$~d, in very good agreement 
with the most recent photometric estimate \citep{Artemenko12}.  

In 2012~Dec (right column of Fig.~\ref{fig:var1}, top panel), RVs of DN~Tau show a more complex 
phase dependence (featuring essentially the second harmonic at half the rotation period) 
and a twice lower amplitude ($\simeq$1.0~\kms) about a mean RV of $\simeq16.9\pm0.1$~\kms.  
This clearly indicates that the cool spot present in 2010~Dec on DN~Tau significantly changed in both 
position and size between our two runs, and shows up in 2012~Dec as a complex of two smaller spots / 
appendages with one appendage on each side of the pole (at phases 0.2 and 0.7 respectively).  

At both epochs, DN~Tau suffers from a moderate amount of spectral veiling of $\simeq$0.5 (at an 
average wavelength of 660~nm).  The veiling is more variable in 2010~Dec than in 2012~Dec, both 
in terms of rotational modulation and intrinsic variability.  In 2010~Dec, maximum veiling (of 
$\simeq$0.7) is reached at phase 0.55, suggesting that the accretion region (presumably causing the 
observed veiling) is close to the cool spot detected from RV variations as in most other cTTSs 
studied to date;  in 2012~Dec, the veiling is more or less constant with phase, suggesting that 
the accretion region is much closer to the pole than during our first run.  

Clear Zeeman signatures are detected at both epochs in the LSD profiles of DN~Tau, usually featuring 
an antisymmetric shape (with respect to the line centre) with a full amplitude of up to 0.8\% (of 
the unpolarized continuum level).  Again, rotational modulation (and intrinsic variability) is 
larger in 2010~Dec, with longitudinal fields \citep[derived from Zeeman signatures with the first 
moment technique, e.g., ][]{Donati97b} varying from --300 to 200~G (with typical error bars of 12 
and 25~G for ESPaDOnS and NARVAL data respectively) on a period (of $6.30\pm0.15$~d) that agrees 
well with the estimate of \citet{Artemenko12}.  

In 2012~Dec, longitudinal fields are both smaller and mostly positive, varying from --40 to 110~G 
(with a median error bar of 11~G), but nevertheless show clear modulation on a period (of 
$6.25\pm0.10$~d) fully compatible with rotation;  from the small amplitude of the rotational 
modulation, we can guess that the magnetic topology (or at least its most intense low-order component) 
is better aligned with the rotation axis in our second observing run.  
As for the veiling, the intrinsic variability of longitudinal fields is smaller in 2012~Dec than 
in 2010~Dec, suggesting that it likely reflects potential effects of unsteady accretion on Zeeman 
signatures rather than rapid changes in the large-scale field topology.  

\subsection{\caii\ IRT emission}

\begin{figure*}
\center{\includegraphics[scale=0.85]{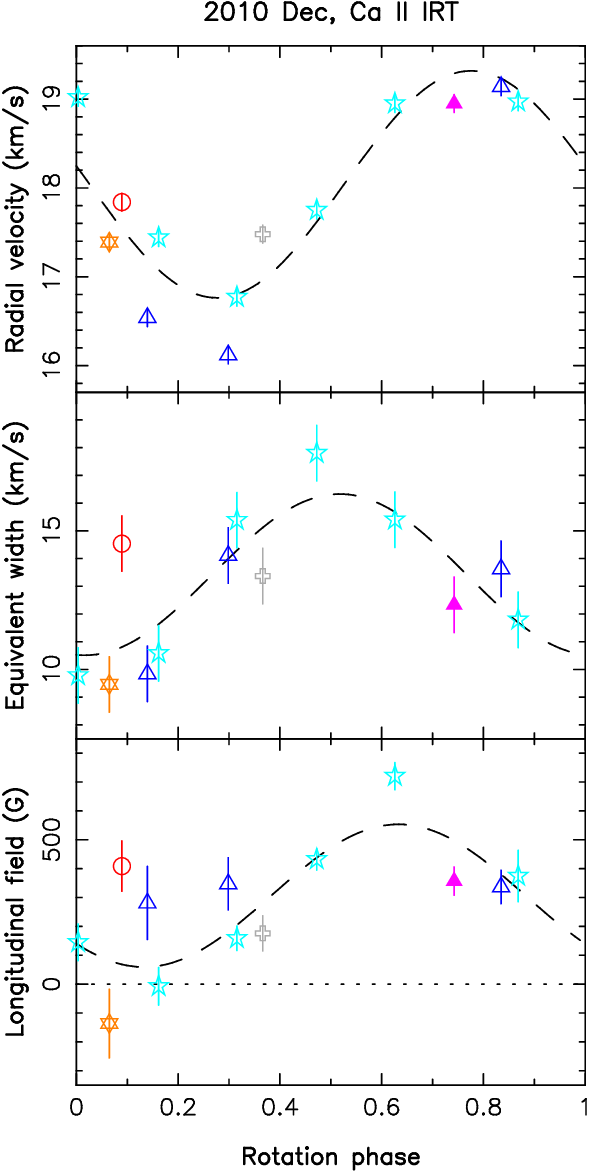}\hspace{4mm}
\includegraphics[scale=0.85]{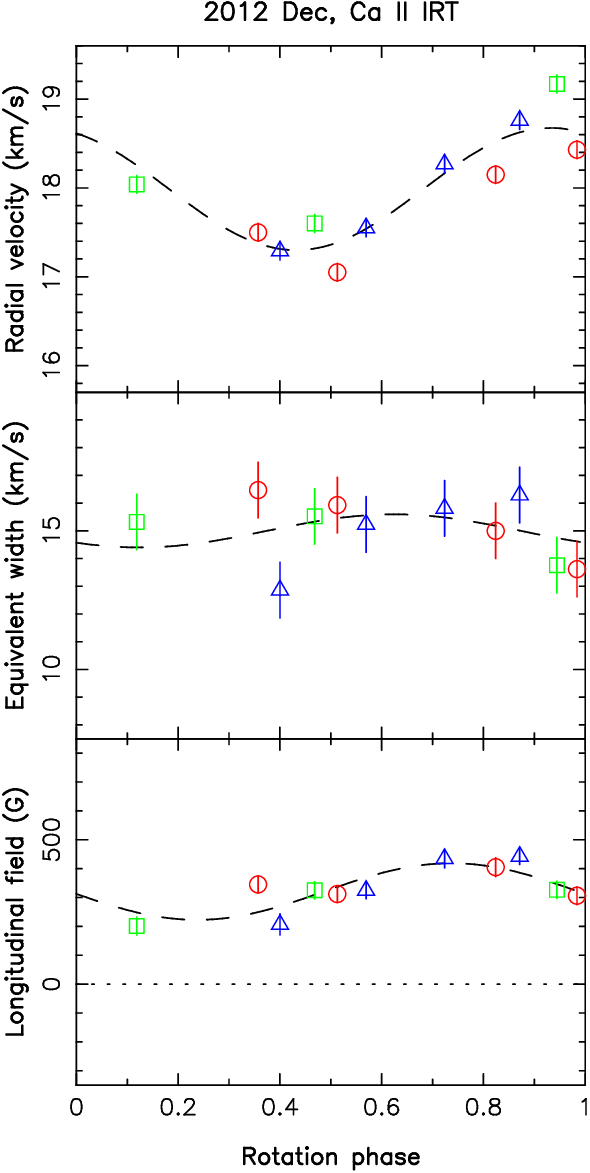}}
\caption[]{RVs (top row), equivalent widths (second row) and longitudinal fields (bottom row)
derived from the \caii\ IRT LSD profiles of DN~Tau in late 2010 (left panels) and late 2012 
(right panels), with symbol/color coding as in Fig.~\ref{fig:var1}.
Conservative error bars of $\pm$0.1 and $\pm$1~\kms\ were assumed for the RVs and equivalent widths
of the emission profile.  This figure is best viewed in color.  }
\label{fig:var2}
\end{figure*}

Narrow emission in the core of \caii\ IRT lines is a very convenient proxy of accretion at the 
surface of low-mass cTTSs, despite tracing simultaneously their non-accreting chromospheres.  
Being formed closer to the stellar surface and in a more static atmosphere than the \hei\ $D_3$ 
emission line (the usual accretion proxy for cTTSs), the narrow emission cores of the \caii\ IRT 
lines feature Zeeman signatures that are simple in shape (almost antisymmetric with respect to the 
line centre) and thus quite easy to model (with no need to account for velocity gradients in the line 
formation region);  moreover, the triplet nature and the IR location of these lines both ensure 
high-data quality, usually overcompensating the signal dilution from the non-accreting chromospheres.  

The accretion proxy we consider here (as in the previous papers of the series) is a LSD-like 
weighted average of the 3 IRT lines, whose unpolarized profile is corrected by subtracting the 
underlying (much wider) Lorentzian absorption profile \citep[with a single Lorentzian fit to the far 
line wings, see, e.g.,][for an illustration]{Donati11b}.  As for photospheric LSD profiles, we 
investigate how the RV, equivalent width and longitudinal field of this composite emission profile 
vary with time in the case of DN ~Tau, throughout each of our observing runs;  the results we 
obtain are presented graphically on Fig.~\ref{fig:var2}.  

We again find that rotational modulation is significantly larger in 2010~Dec (than in 2012~Dec).  
At this epoch, RVs of the core emission profile are found to vary by up to 3~\kms\ 
peak-to-peak, in almost perfect anti-phase with those of LSD profiles and with a period 
(of $6.22\pm0.10$~d) in good agreement with that used to phase our data.  It suggests 
that the parent accretion region is centred at phase 0.50 (i.e., at mid-phase between RV 
minimum and maximum);  this is also the phase at which core emission in \caii\ lines 
(varying from about 9 to 18~\kms, or equivalently from 0.027 to 0.050~nm) 
reaches maximum strength (see Fig.~\ref{fig:var2}, middle left panel) and at which 
veiling is maximum (see Fig.~\ref{fig:var1}, middle left panel), further 
confirming our conclusion that the accretion region is more or less coincident with the 
cool spot traced by photospheric LSD profiles.  

In 2012~Dec, the RV curve suggests that the accretion region is centred at phase 0.70 
and located closer to the pole (with a peak-to-peak amplitude of only 2~\kms).  This is 
fully compatible with the reported emission strengths in the core of the \caii\ IRT lines
(exhibiting a much lower level of variability, about a mean of 15~\kms\ or 0.043~nm, see 
middle right panel of Fig.~\ref{fig:var2}) and with the more-or-less constant veiling detected 
at this epoch.  We further infer that this accretion 
region is simpler and less extended than the cool spot traced by photospheric LSD profiles 
(whose complex shape was readily visible from the corresponding RV curve, see 
Fig.~\ref{fig:var1}, top right panel).  

\begin{figure*}
\center{
\includegraphics[scale=0.65,angle=-90]{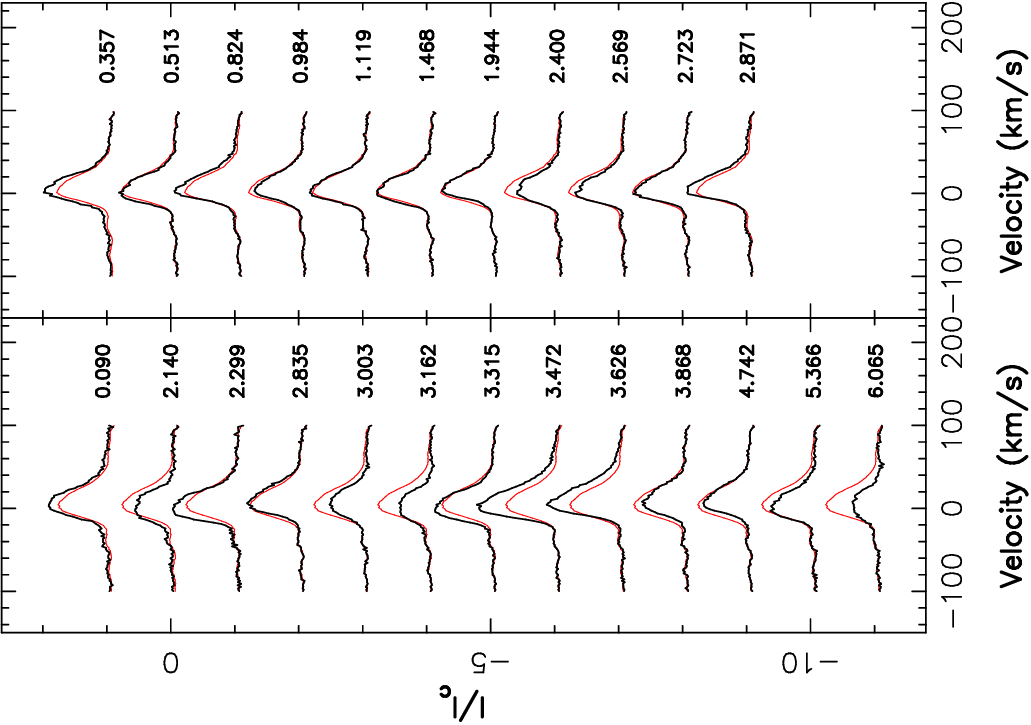}\hspace{5mm}
\includegraphics[scale=0.65,angle=-90]{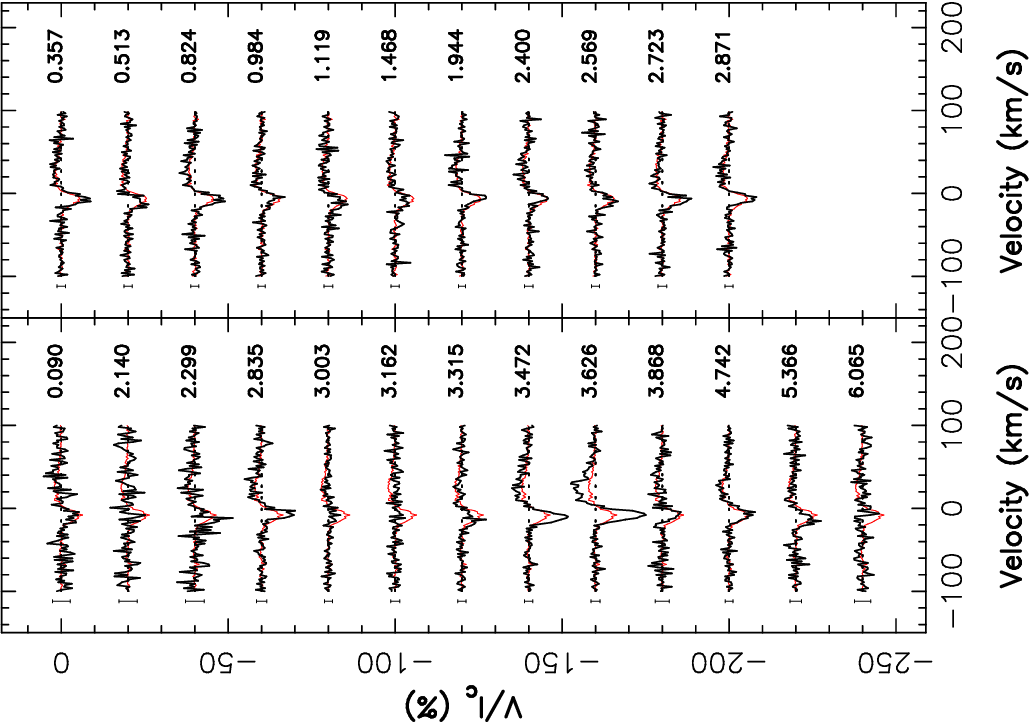}}
\caption[]{Variations of the unpolarized (Stokes $I$, left panel) and circularly-polarized
(Stokes $V$, right panel) profiles of the \hei\ $D_3$ emission of DN~Tau in 2010~Dec (left
columns of both panels) and 2012~Dec (right columns).
Clear Zeeman signatures (with full amplitudes of up to 20\% in 2009~July)
are detected at most epochs;  these signatures obviously feature a stronger / narrower blue 
(negative) lobe and a weaker / wider red (positive) lobe, i.e., have shapes that clearly depart 
from the usual antisymmetric pattern with respect to the line centre (i.e., with similarly strong / wide 
blue and red lobes).  To emphasize variability, the average profile over each run is shown in red.
Rotation cycles (as listed in Table~1) and 3$\sigma$ error bars (for Stokes $V$ profiles only)
are shown next to each profile.  }
\label{fig:he}
\end{figure*}

We note that at both epochs, \caii\ IRT emission cores of DN~Tau are located at an 
average RV of $\simeq$18~\kms, i.e., red-shifted by $\simeq$1~\kms\ with respect to 
photospheric lines, typical to the moderately accreting cTTSs observed so far \citep[e.g.,][]{Donati12}

As for LSD profiles, Zeeman signatures are clearly detected in association with the narrow emission 
cores of the \caii\ IRT lines, with a full amplitude of up to 10\% peak to peak (in units of the 
unpolarized continuum).  The corresponding longitudinal fields range from --140 to 720~G in 2010 
and from 200 to 440~G in 2012 (with typical error bars of 40 and 100~G for ESPaDOnS and NARVAL data respectively), 
i.e., exhibiting both a larger rotational modulation (with a period of 
$6.19\pm0.20$~d) and intrinsic variability in the first of our observing runs (see lower panels of Fig.~\ref{fig:var2}).  
Maximum longitudinal fields are reached around phase 0.60 and 0.70 in 2010 and 2012 respectively, suggesting that 
the magnetic pole is more or less coincident with the cool polar spot and accretion region already tracked 
through the other spectral diagnostics (RVs and equivalent widths);  the much smaller rotational modulation 
observed in 2012 also suggests the large-scale field is variable on a timescale of 2~yr, being better aligned 
with the rotation axis in 2012 than in 2010.  

We note that longitudinal fields of DN~Tau as derived from LSD profiles and \caii\ emission cores are not systematically 
of opposite sign, as for other previously studied cTTSs \citep[e.g., TW~Hya,][]{Donati11b}, being 
even both mainly of the same sign most of the time in 2012 \citep[as for AA~Tau, e.g.,][]{Donati10b};  
moreover, longitudinal field strengths in LSD profiles are, once averaged over the rotation cycle, much lower than 
those in \caii\ profiles (by typically a factor of 5 or more) rather than being comparable as for TW~Hya.  
This suggests that the large-scale field at the surface of DN~Tau is globally simpler than that of TW~Hya, 
featuring in particular a weaker degree of polarity reversals over the visible hemisphere, and shares 
similarities with that of AA~Tau.  

\subsection{\hei\ $D_3$ emission}

\begin{figure*}
\center{\includegraphics[scale=0.85]{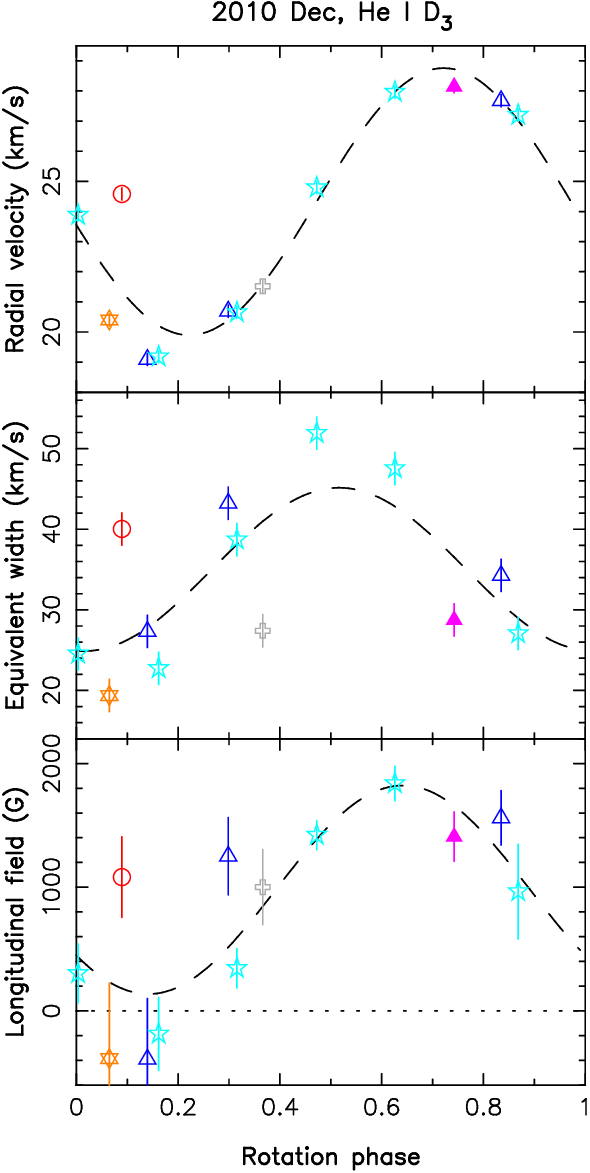}\hspace{4mm}
\includegraphics[scale=0.85]{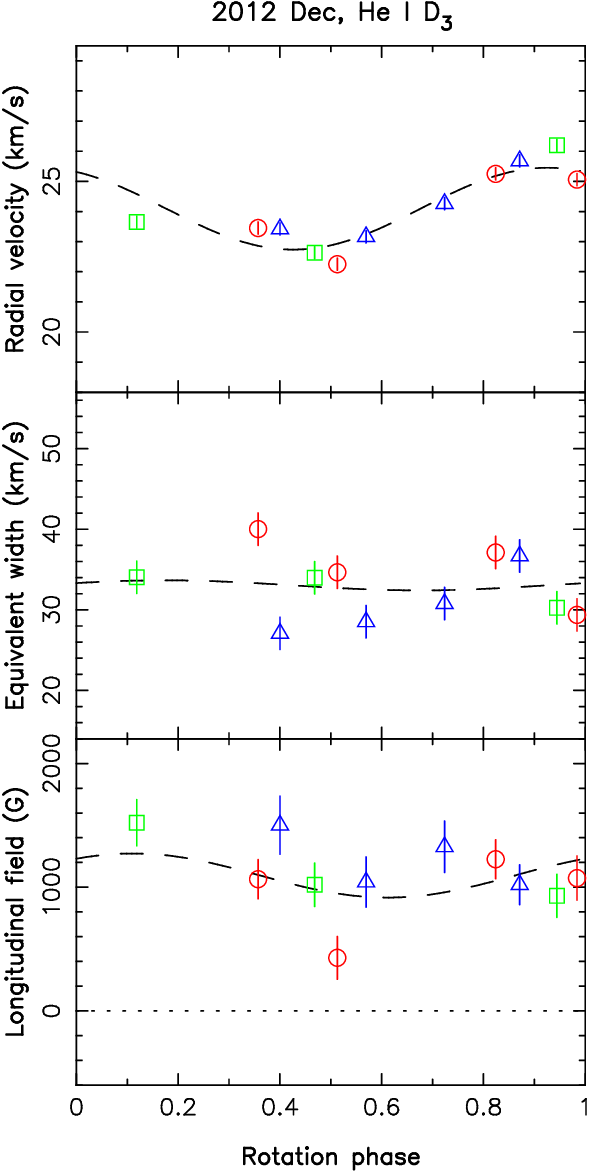}}
\caption[]{Same as Fig.~\ref{fig:var2} for the narrow component of the \hei\ $D_3$ profiles of DN~Tau.
Conservative error bars of $\pm$0.2~\kms\ and $\pm$2~\kms\ were assumed on the RVs and equivalent widths
of the emission profile.  This figure is best viewed in color.  }
\label{fig:var3}
\end{figure*}

The unpolarized and circularly-polarized line profiles of the more conventional accretion proxy often used 
in cTTSs studies, i.e., the \hei\ $D_3$ line, are shown in Fig.~\ref{fig:he}, as well as their variations with 
time (with respect to the average) for our two observing runs.  They both feature a clearly asymmetric shape, 
with Stokes $I$ profiles exhibiting a broader red wing (compared to the blue one) and Stokes $V$ profiles 
showing a blue lobe that is deeper and narrower (than the red one).  These asymmetries directly reflect that 
the \hei\ $D_3$ emission of cTTSs is formed in a region featuring strong velocity gradients, presumably probing the 
postshock zone of the accretion region where the incoming disc plasma is experiencing a strongly decelerating fall 
towards the stellar surface.  We also note that the \hei\ $D_3$ 
line of DN~Tau at both epochs only shows a relatively narrow emission component, but no broad emission component 
as in a number of cTTSs \citep[e.g., TW~Hya or GQ~Lup][]{Donati11b, Donati12}.  

Rotational modulation of the RVs, equivalent widths and longitudinal fields derived from the \hei\ emission, 
shown in Fig.~\ref{fig:var3}, essentially repeat those inferred from the \caii\ IRT emission core.  
In 2010, RVs vary more or less sinusoidally with phase, with a period of $6.17\pm0.10$~d (again 
compatible with that used to phase the data) and a semi-amplitude of $\simeq$4.5~\kms\ 
(about half the \vsini\ of DN~Tau, see Sec.~\ref{sec:mod}).  Assuming as usual that most of the \hei\ emission 
comes from the accretion region, it indicates that this region is located at phase 0.50 (i.e., at mid time between 
RV minimum and maximum, fully compatible with results from \caii\ IRT emission) and at latitude $\simeq$60\degr.  
We also further confirm that RVs vary much less in 2012 (full amplitude of $\simeq$4~\kms), indicating that the 
accretion region has moved much closer to pole, at latitude $\simeq$75\degr\ and phase 0.70.  
In both cases, \hei\ emission occurs at an average RV of $\simeq$24~\kms, i.e., red-shifted by $\simeq$7~\kms\ with 
respect to average LSD photospheric profiles;  as with other cTTSs \citep[e.g., GQ~Lup,][]{Donati12}, this 
confirms that \hei\ emission is being formed in non-static atmospheric layers, as already evidenced by the strong 
asymmetries in both Stokes $I$ and $V$ \hei\ profiles (see Fig.~\ref{fig:he}).  

Equivalent widths of \hei\ emission ranges from about 20 to 50~\kms\ (i.e., 0.040 and 0.100~nm) with an average 
value of $\simeq$35--40~\kms\ (i.e., 0.070--0.080~nm, see Fig.~\ref{fig:var3} mid panels).  In 2010, maximum 
emission is reached at phase 0.50, i.e., when the accretion region (as derived from RV curves) is best viewed from 
the observer, as expected.  In 2012, \hei\ emission is more or less constant with time, in agreement with our finding (from the RV curve) 
that the accretion region is located much closer to the pole at this epoch.  Again, rotational modulation 
of the equivalent width of \hei\ emission mostly repeats that derived from \caii\ IRT lines.  

\begin{figure*}
\center{
\includegraphics[scale=0.65,angle=-90]{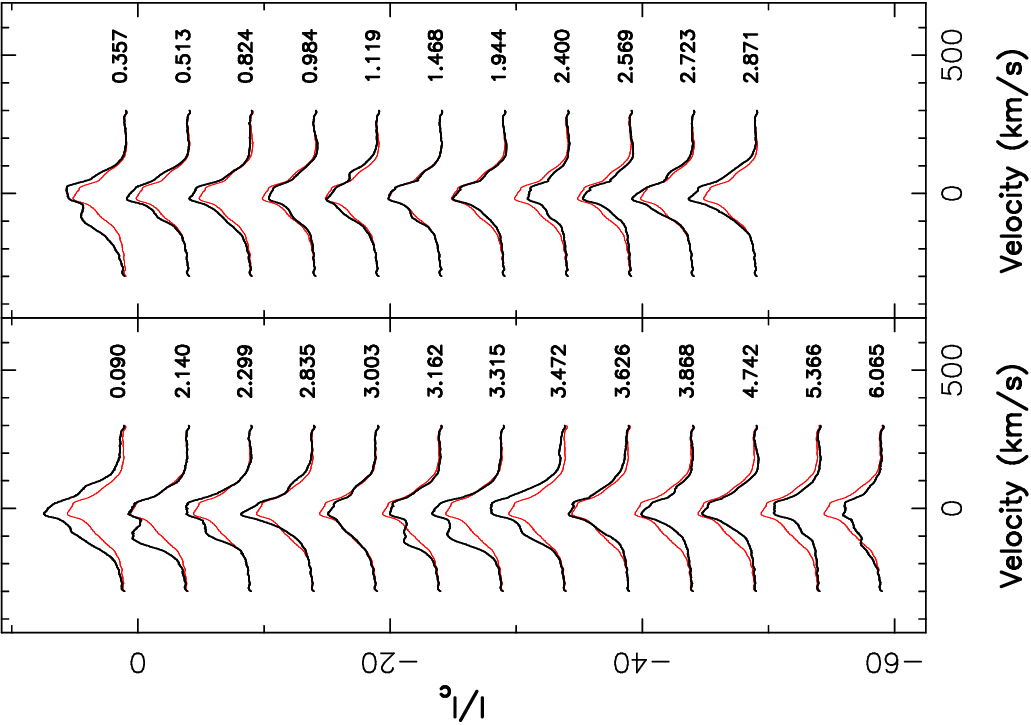}\hspace{5mm}
\includegraphics[scale=0.65,angle=-90]{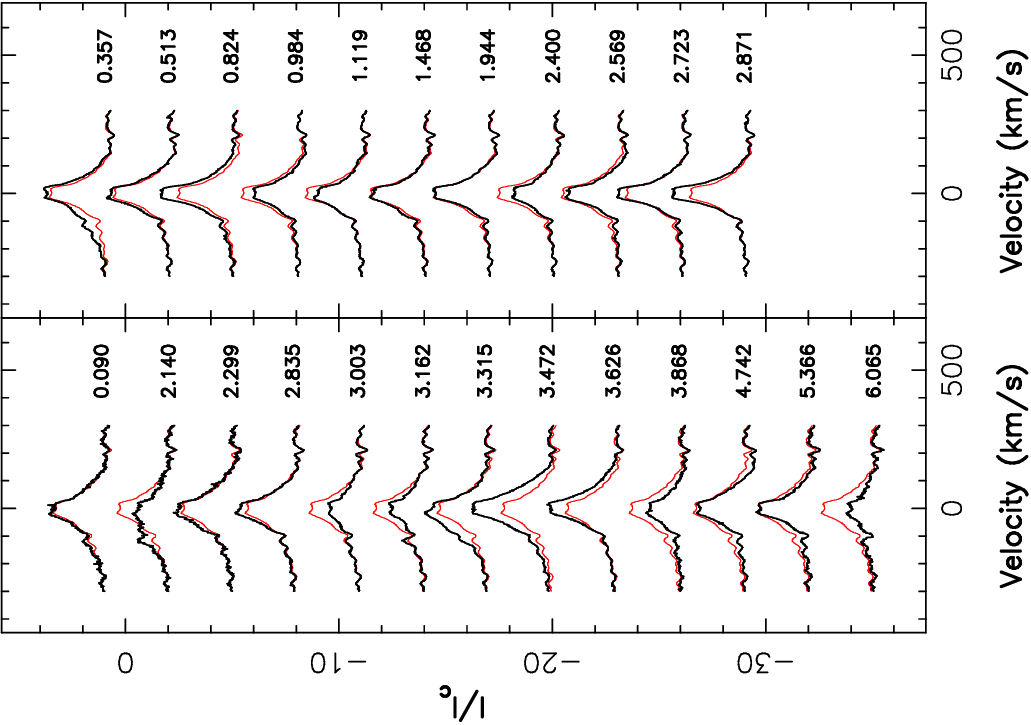}}
\caption[]{Variations of the \hal\ (left) and \hbe\ (right) lines in the spectrum of
DN~Tau, in 2010~Dec (left column of both panels) and 2012~Dec (right column).
To emphasize variability, the average profile over each run is shown in red.
Rotation cycles (as listed in Table~1) are mentioned next to each profile.  }
\label{fig:bal}
\end{figure*}

Zeeman signatures are detected at most epochs in conjunction with \hei\ emission, indicating the presence of 
longitudinal fields of up to 1.8~kG in the accretion region of DN~Tau (see Fig.~\ref{fig:var3} bottom panels).  
Rotational modulation of Zeeman signatures is clear in 2010, with longitudinal fields varying from --0.4 to 1.8~kG 
(typical error bars of about 200~G for ESPaDOnS data and 400~G for NARVAL data) with a period of $6.10\pm0.15$~d, 
in agreement with a magnetic pole significantly offset from the rotation pole.  We also note that, as for \caii\ 
emission, longitudinal fields peak at a slightly later phase (of about 0.60) than equivalent widths, suggesting that 
the accretion spot is close but not exactly coincident with the magnetic pole.  In 2012, longitudinal fields are 
roughly constant (at about 1.2~kG), the observed temporal evolution being mainly caused by intrinsic variability 
rather than rotational modulation;  this further confirms that the large-scale field of DN~Tau is evolving on a 
timescale of 2~yr, reaching in 2012 a state of near-alignment with the rotation axis.  

\subsection{Balmer emission}

Balmer \hal\ and \hbe\ emission profiles of DN~Tau are shown in Fig.~\ref{fig:bal} at both epochs.  For most epochs during our runs, 
these profiles consist of a central emission peak with an average equivalent width of about 700 and 300~\kms\ (i.e., 1.5 and 0.50~nm) 
respectively;  additional blue emission (shifted by about 100~\kms\ with respect to the main central emission) also shows up 
sporadically in both profiles, though much more conspicuously in \hal\ than in \hbe\ (e.g., at cycles 2.140 and 3.315 in 2010, 
or at cycle 0.357 in 2012, see Fig.~\ref{fig:bal}).  

Looking at variance profiles (not shown here), we obtain that variability occurs in 3 separate regions of these Balmer lines.  
Most of the variability takes place either in the central emission peak or in the additional blue emission mentioned above, the 
strongest variability occurring in the central emission peak for \hbe\ and in the blue emission component for \hal.  When 
most conspicuous (in 2010), the blue \hal\ emission component is found to vary in a more or less anti-correlated way with the 
main central emission.  This variability, and in particular the one associated with the blue emission component, is apparently 
not related to rotational modulation, rather discrepant profiles being observed at similar phases of different rotation cycles 
(e.g., cycles 2.140 and 3.162, or 3.315 and 5.365 in 2010, or 0.357 and 2.400 in 2012).  

Variability is also observed, though 
at a much weaker level, in the red wing of both Balmer profiles (at velocities of 100--200~\kms);  for contrast reasons, 
this variability is better observed in \hbe\ and shows up as weak absorption transients only visible in a small number of 
spectra (e.g., at cycle 4.742 in 2010, or 2.569 in 2012).  This variability is somewhat reminiscent of the strong 
absorption features occurring in the red wing of Balmer lines of AA~Tau (attributed to the recurring 
transits of a magnetically-confined accretion funnel over the visible hemisphere of this mainly edge-on prototypical cTTS);  
however, these red absorption episodes are much weaker in DN~Tau (possibly as a result of the low inclination) 
and possibly less systematic as well (no clear absorption being observed at similar phases of different rotation cycles, 
e.g., at phase 2.835 in 2010).  

\begin{figure*}
\center{\hbox{
\includegraphics[scale=0.37,angle=-90]{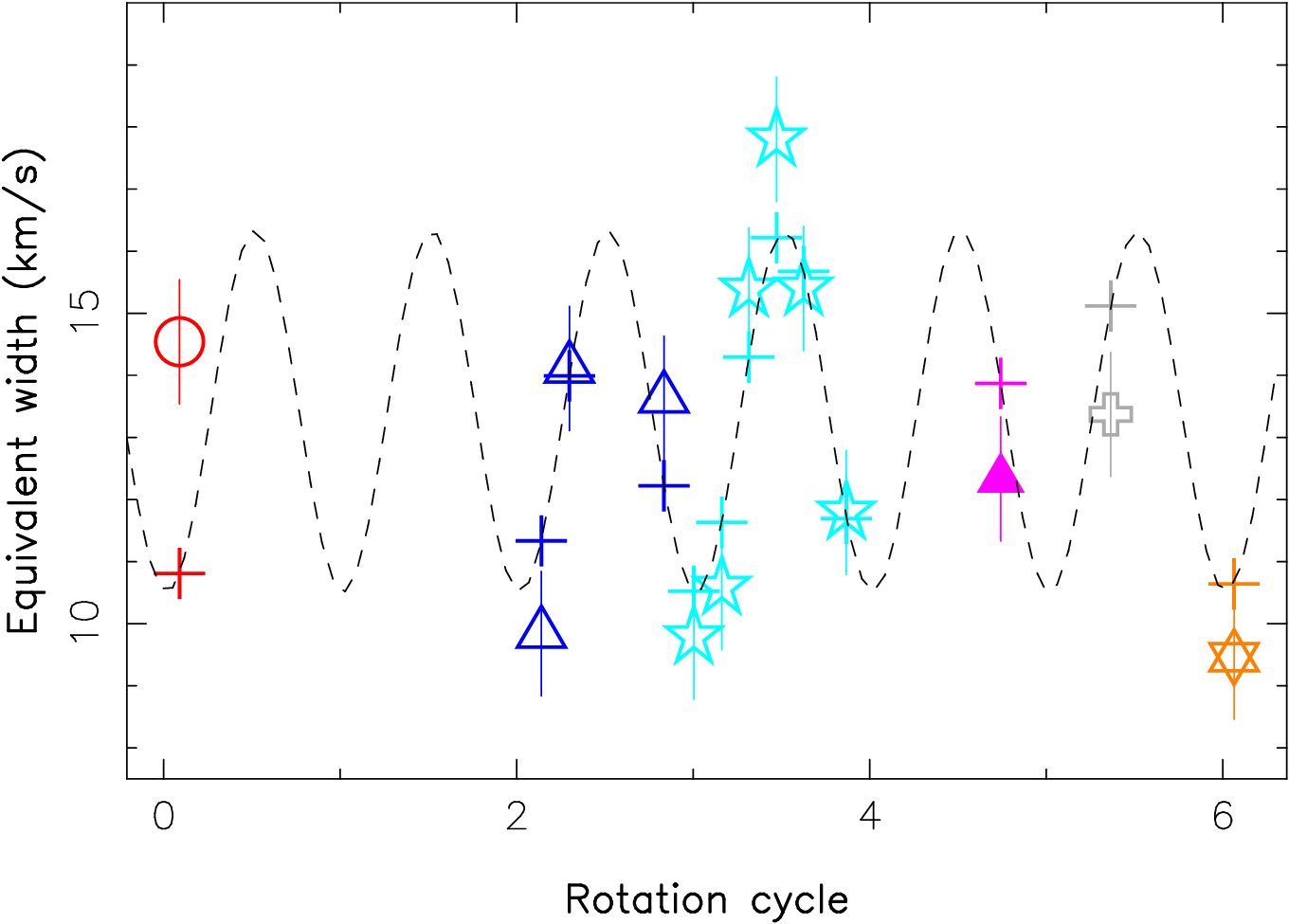}\hspace{5mm}
\includegraphics[scale=0.37,angle=-90]{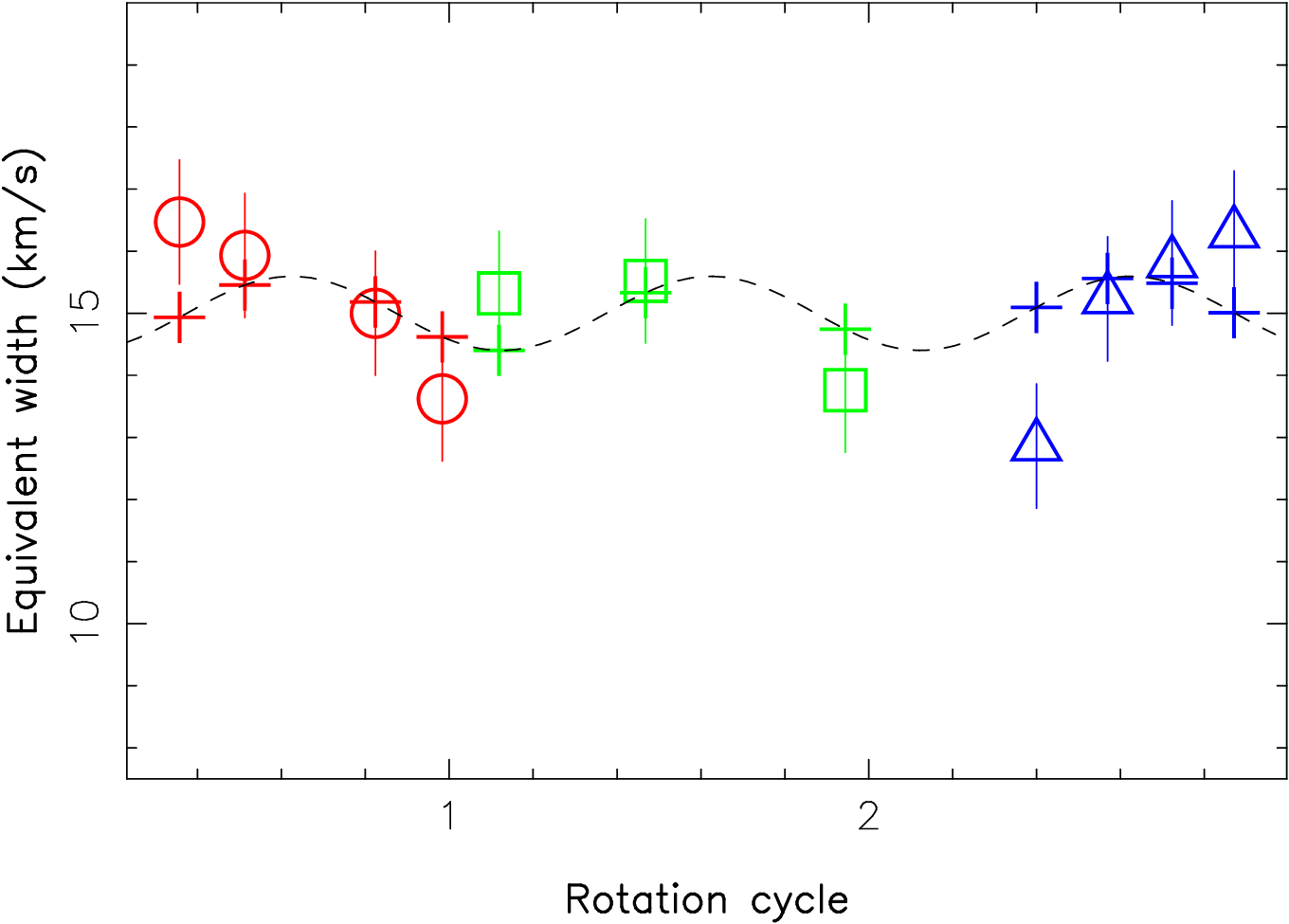}}}
\caption[]{Measured (open symbols) and fitted (pluses) equivalent widths of
the \caii\ IRT LSD profiles of DN~Tau~ in 2010~Dec (left panel) and 2012~Dec (right panel).
The model wave (dashed line) providing the best (sine+cosine) fit to
the data presumably traces rotational modulation (with a period of 6.32~d), while the deviation
from the fit illustrates the level of intrinsic variability.  
The open symbols are defined as described in Fig.~\ref{fig:var1}.  This figure is best viewed in color.  }
\label{fig:ew}
\end{figure*}

\subsection{Mass-accretion rate}

From the emission strength of the various accretion proxies mentioned above, we can derive an estimate of the mass accretion rate 
using the same method as that outlined in the previous papers of the series \citep[e.g.,][]{Donati12}.  Approximating the stellar 
continuum by a Planck function at temperature 3,950~K, we derive, from the \caii\ IRT, \hei, \hbe\ and \hal\ equivalent widths 
reported above (respectively equal to 0.045, 0.075, 0.50 and 1.5~nm), logarithmic line fluxes (in units of the solar luminosity 
\lsun) of --3.5, --4.2, --4.9 and --5.1.  Using the empirical calibrations of \citet{Fang09}, these line fluxes translate into 
logarithmic accretion luminosities of --2.2, --2.4, --1.7 and --2.3 respectively (again in units of \lsun).  From this we infer 
that the average logarithmic mass accretion rate at the surface of DN~Tau (in units of \mspy) is $-9.1\pm0.3$ at both epochs 
(with estimates derived from individual proxies agreeing all with this value within the quoted error bar).  
Very similar results are obtained (logarithmic mass accretion rate of $-9.2\pm0.3$~\mspy) when using the newer empirical 
calibrations of \citet{Rigliaco12}.  Our estimate is also in rough agreement with the older one of \citet{Gullbring98}, 
equal to $-8.5$~\mspy, based on measurements of the UV continuum excess presumably produced by accretion - especially 
if we allow for likely systematic differences between both types of measurements \citep[see, e.g.,][]{Rigliaco12}, as well as 
for potential long-term variations of the accretion rate between our observations and those used by \citet{Gullbring98}.  

An independent (though much less accurate) estimate can also be obtained from the width of \hal\ \citep[e.g.,][]{Natta04}.  
Given the full width at 10\% height of $300\pm20$~\kms\ of the average \hal\ profile of DN~Tau, we infer a logarithmic mass 
accretion rate of $-10.0\pm0.6$, smaller and marginally compatible with our previous estimate.  However, as already stressed by 
\citet{Cieza10}, the precision of this measurement technique depends in an unclear way on the overall shape of \hal\ (and of 
its variations with time), rendering this accretion proxy a relatively poor quantitative indicator in practice;  for this reason, 
we disregard this latter estimate and only consider our former (and main) measurement of the accretion rate in the remaining 
sections of our paper.

\section{Magnetic modelling}
\label{sec:mod}

After this qualitative overview and preliminary analysis of our data sets, we now propose a more quantitative and 
unbiased modeling of the photospheric LSD and \caii\ IRT unpolarized and circularly-polarized profiles 
in terms of maps of the large-scale magnetic topology, and of distributions of photospheric cool spots and 
of chromospheric accretion regions, at the surface of DN~Tau.  The software tool we are using in this aim is a 
dedicated stellar-surface tomographic-imaging package based on the principles of maximum-entropy image reconstruction 
and on the main assumption that the observed variability is mainly caused by rotational modulation;  our code was 
adapted to the specific needs of MaPP observations \citep{Donati10b} and extensively tested on a number of previous 
data sets \citep[e.g.,][]{Donati11, Donati12} including close binary stars \citep{Donati11c}.  

More specifically, the code is set up to invert (both automatically and simultaneously) time series of Stokes 
$I$ and $V$ LSD and \caii\ profiles into magnetic, brightness and accretion maps of the observed protostar.  
The reader is referred to \citet{Donati10b} for more details on the imaging method.  

\subsection{Application to DN~Tau}

We start the process by applying to our data sets the usual filtering techniques, designed 
for retaining rotational modulation while discarding intrinsic variability \citep[see, e.g.,][for more 
information]{Donati10b, Donati12}.  The effect of this process on \caii\ profiles is illustrated in 
Fig.~\ref{fig:ew} in the particular case of DN~Tau.  
In practice, this filtering has little impact on the reconstructed maps, and essentially allows us to 
ease the convergence of the iterative optimization algorithm on which the imaging code is based.  

The local Stokes $I$ and $V$ profiles are synthesized using Unno-Rachkovsky's solution to the equations 
of polarized radiative transfer in a Milne-Eddington model atmosphere, known to provide a reliable 
description (including magneto-optical effects) of how shapes of line profiles are distorted in the 
presence of magnetic fields \citep[e.g.,][]{Landi04}.  The main local line parameters used for DN~Tau 
are again very similar to those used in our previous studies.  For the average photospheric LSD profile, 
the wavelength, Doppler width, unveiled equivalent width and Land\'e factor are respectively set to 
660~nm, 1.9~\kms, 4.2~\kms\ and 1.2;  for the quiet \caii\ emission profile, they are respectively set 
to 850~nm, 7~\kms, 10~\kms\ and 1.0.  Finally, we assume that \caii\ emission is locally enhanced in 
accretion regions (with respect to quiet chromospheric regions) by a factor $\epsilon=10$, as in all 
previous studies.  

\begin{figure*}
\vspace{-2mm}
\hbox{\includegraphics[scale=0.7]{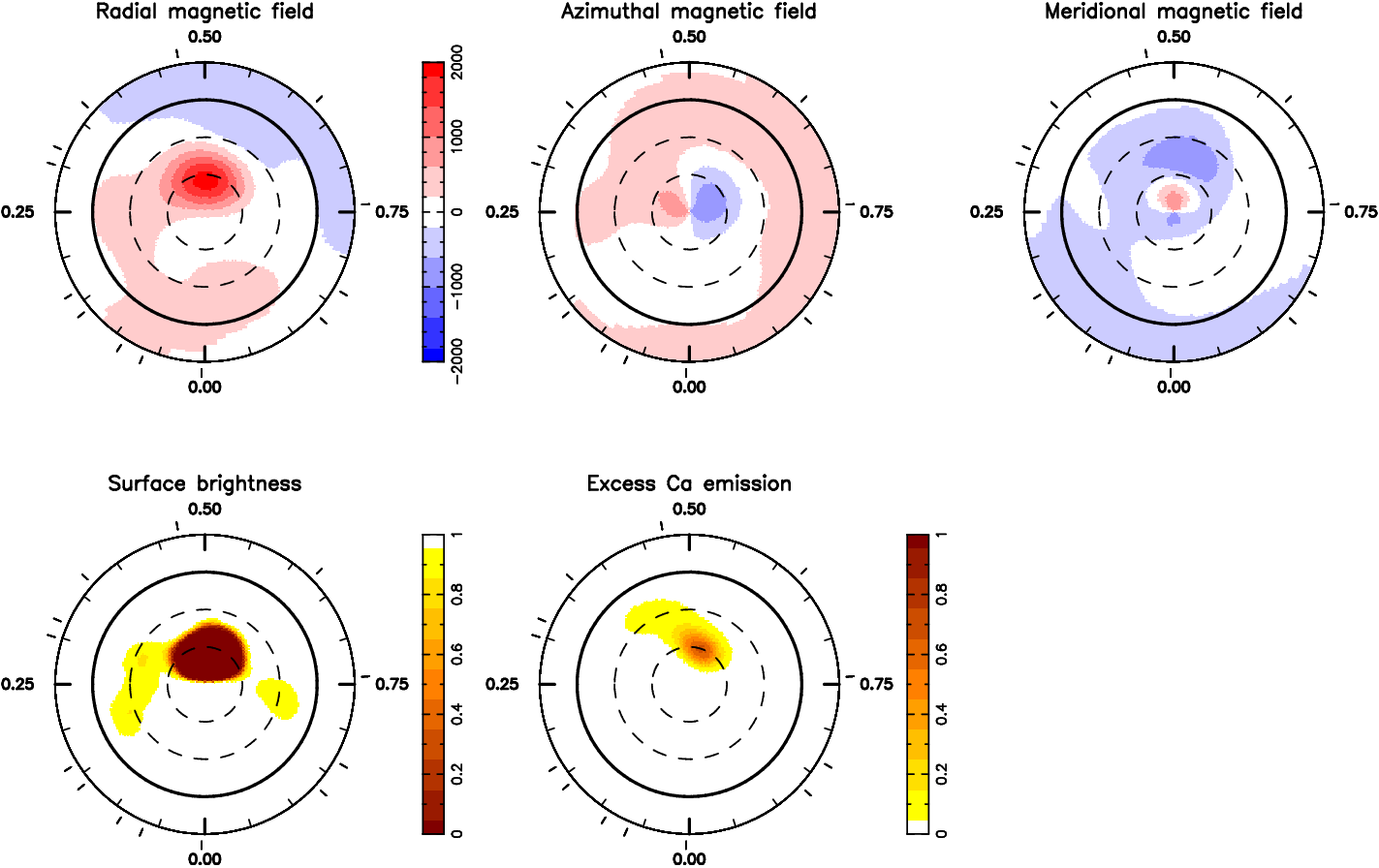}}
\vspace{8mm}
\hbox{\includegraphics[scale=0.7]{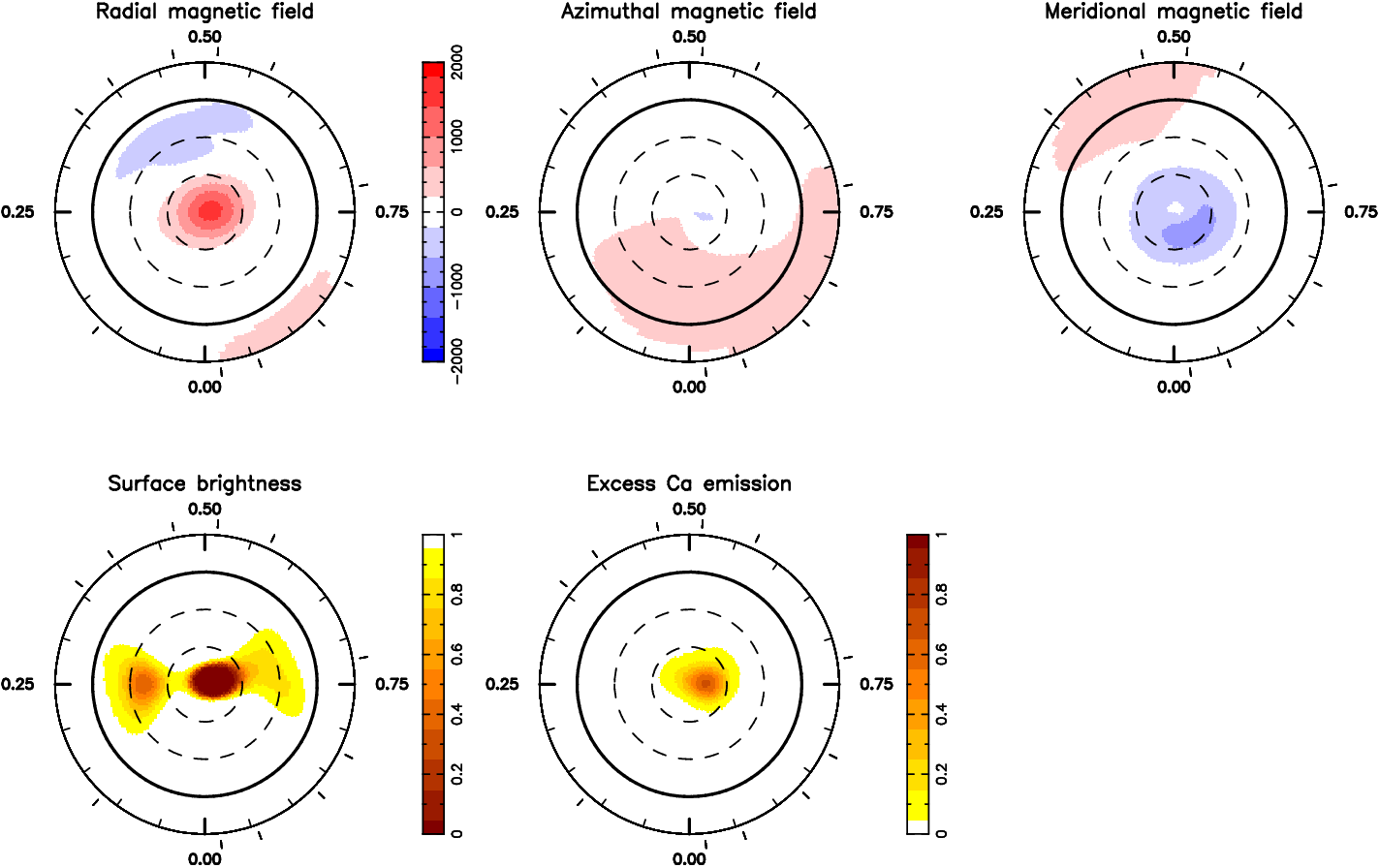}}
\caption[]{Maps of the radial, azimuthal and meridional components of the magnetic field $\bf B$
(first and third rows, left to right panels respectively), photospheric brightness and excess
\caii\ IRT emission (second and fourth rows, first and second panels respectively) at the
surface of DN~Tau, in 2010~Dec (top two rows) and 2012~Dec (bottom two rows).
Magnetic fluxes are labelled in G;  local photospheric brightness (normalized to that of the quiet
photosphere) varies from 1 (no spot) to 0 (no light);  local excess \caii\ emission varies from 0
(no excess emission) to 1 (excess emission covering 100\% of the local grid cell, assuming an
intrinsic excess emission of 10$\times$ the quiet chromospheric emission).
In all panels, the star is shown in flattened polar projection down to latitudes of $-30\degr$,
with the equator depicted as a bold circle and parallels as dashed circles.  Radial ticks around
each plot indicate phases of observations.  This figure is best viewed in color. }
\label{fig:map}
\end{figure*}

\begin{figure*}
\vspace{-3mm}
\center{
\includegraphics[scale=0.65,angle=-90]{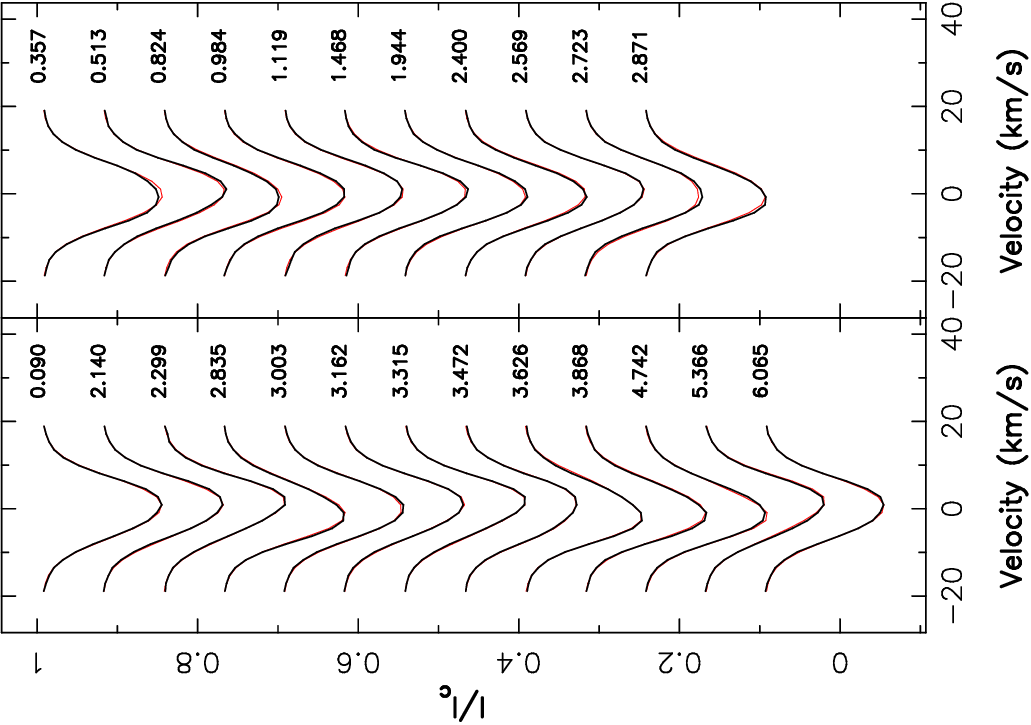}\hspace{4mm}
\includegraphics[scale=0.65,angle=-90]{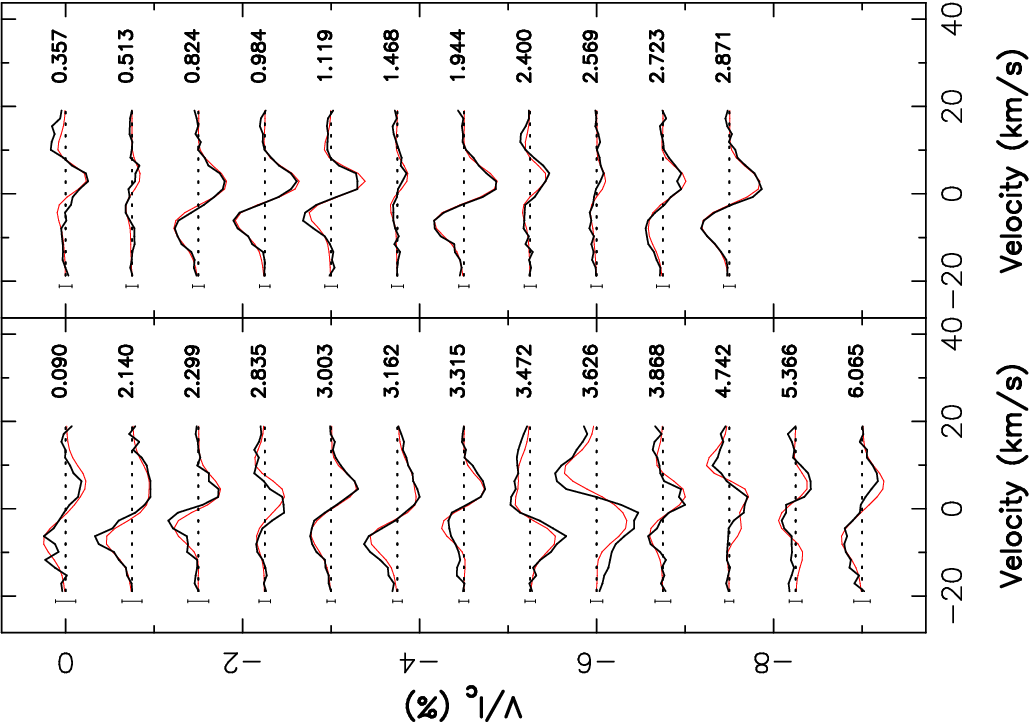}}
\vspace{3mm}
\center{
\includegraphics[scale=0.65,angle=-90]{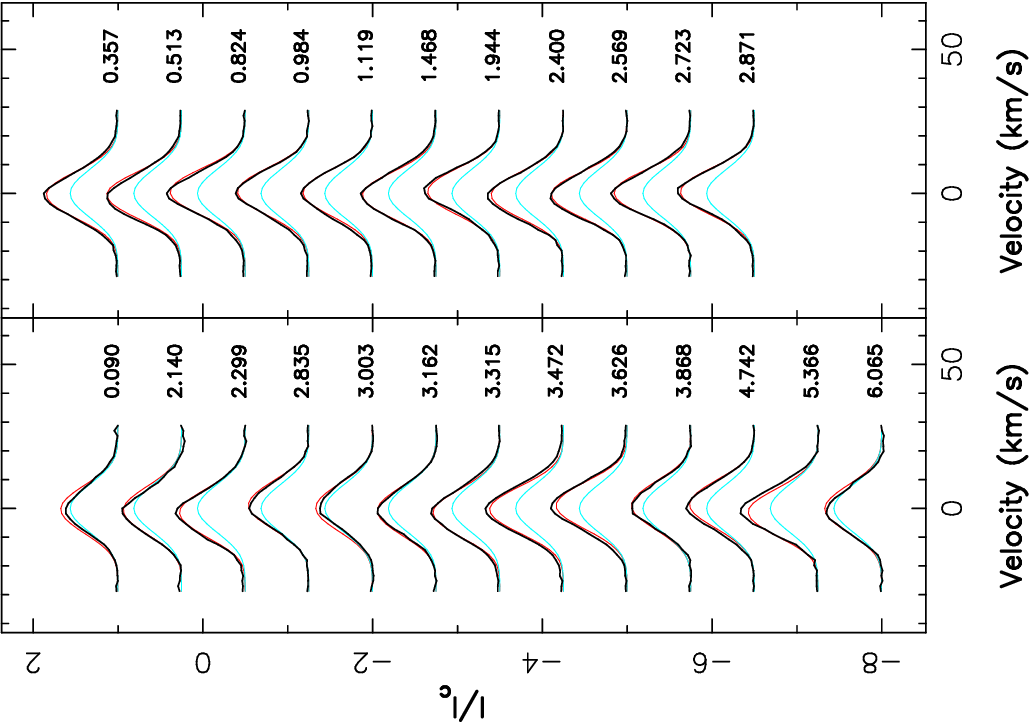}\hspace{4mm}
\includegraphics[scale=0.65,angle=-90]{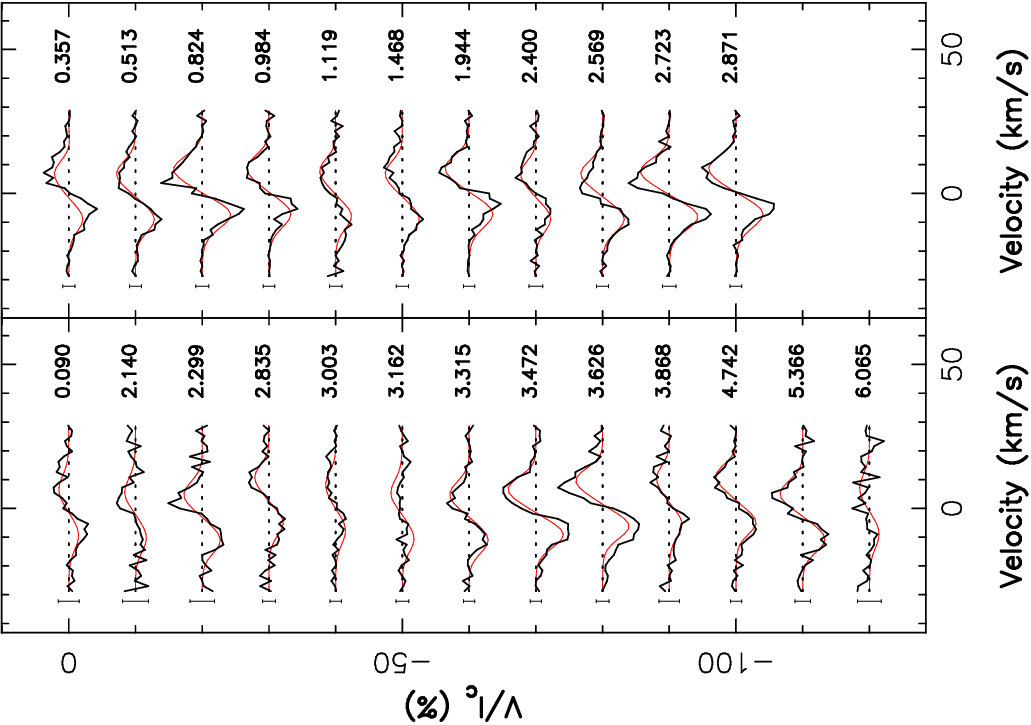}}
\caption[]{Maximum-entropy fit (thin red line) to the observed (thick black line) Stokes $I$ and
Stokes $V$ LSD photospheric profiles (top panels) and \caii\ IRT profiles (bottom panels)
of DN~Tau.  In each panel, the left and right columns correspond to the 2010~Dec and 2012~Dec
data respectively.  The light-blue curve in the bottom left panel shows the (constant)
contribution of the quiet chromosphere to the Stokes $I$ \caii\ profiles.
Rotational cycles and 3$\sigma$ error bars (for Stokes $V$ profiles) are also shown next to each
profile.  }
\label{fig:fit}
\end{figure*}

The reconstructed magnetic, brightness and accretion maps of DN~Tau are shown in Fig.~\ref{fig:map} 
for both epochs, with the corresponding fits to the data shown in Fig.~\ref{fig:fit}.  Once again, 
we assume that the magnetic topology of DN~Tau is antisymmetric with respect to the centre of the 
star;  the spherical harmonic (SH) expansions describing the reconstructed field is limited to terms
with $\ell\leq7$, which is found to be adequate when $\vsini<10$~\kms.  

Error bars on Zeeman signatures were artificially expanded by a factor of 1.5 (in 2010) to 2 (in 2012), 
both for LSD profiles and for \caii\ emission, to compensate for the level of intrinsic variability 
specific to the Stokes $V$ data, obvious from the lower panels of Figs.~\ref{fig:var1} and \ref{fig:var2} 
(where the observed dispersion on longitudinal fields is larger than formal error bars) and not removed 
by our filtering process (correcting irregular changes in line equivalent widths only).  
The fits we finally obtain correspond to a reduced chi-square \chisqr\ equal to 1, starting
from initial values of about 9 and 16 in 2010 and 2012 respectively (for a null magnetic field and unspotted
brightness and accretion maps, and with scaled-up error bars on Zeeman signatures).

As part of the imaging process, we obtain new, more accurate estimates for a number of parameters of DN~Tau.  
We find in particular that the average RV of DN~Tau is $17.0\pm0.1$~\kms\ at both epochs and that \vsini\ 
is equal to $9\pm1$~\kms, compatible with previous literature estimates \citep[e.g.,][]{Appenzeller05}.  
We also find that Zeeman signatures are best fitted for values of the local filling factor \citep[describing 
the relative proportion of magnetic areas at any given cell of the stellar surface, see][]{Donati10b} equal 
to $\psi\simeq$1, significantly larger than that derived for most cTTSs analysed to date \citep[for which 
$\psi\simeq0.4$, e.g.,][]{Donati11}.  

\subsection{Modelling results}

As clear from Fig.~\ref{fig:map}, the large-scale fields reconstructed for DN~Tau at both epochs are mainly 
axisymmetric.  Their main feature is a region of positive radial field located at a latitude of 60\degr\  
and phase 0.50 in 2010 and almost exactly on the pole in 2012;  the magnetic flux in this region reaches 
about 1.8~kG in 2010 and 1.3~kG in 2012, in good agreement with the maximum longitudinal field values probed 
by the \hei\ accretion proxy (see Fig.~\ref{fig:var3}, lower panels).  Both maps include low-latitude arcs 
of negative radial field resembling those found on previously studied cTTSs \citep[like GQ~Lup,][]{Donati12} 
but with a lower contrast, i.e., a weaker flux relative to that of the main radial field region;  similarly, 
both maps also feature a high-latitude ring of mostly negative (i.e., equatorward) meridional field 
surrounding the main magnetic pole.  

The reconstructed field is mostly poloidal, the toroidal component storing no more than 10--15\% of the 
overall magnetic energy.  The poloidal component is mostly axisymmetric in 2012, with more than 80\% of
the magnetic energy concentrating in SH modes with $m<\ell/2$ ($\ell$ and $m$ denoting respectively 
the degrees and orders of the modes);  this fraction drops down to only $\simeq$50\% in 2010 as a result 
of the significant tilt of the large-scale field (with respect to the rotation axis).  

We find that the octupole field dominates the large-scale surface field at both epochs, concentrating 
about 45 and 60\% of the poloidal magnetic energy in 2010 and 2012 respectively, whereas the dipole field 
only stores 35 and 20\% of the energy at these 2 epochs;  their corresponding strengths at the surface 
of DN~Tau are respectively equal to 780 and 600~G in 2010 and 2012 for the octupole component, and to 
530 and 300~G for the dipole component, implying octupole to dipole intensity ratios of 1.5 and 2.0 
at both epochs.  This confirms in particular that the large-scale field of DN~Tau significantly weakened 
between 2010 and 2012, as guessed from the long-term evolution of the longitudinal field curves (see 
Sec.~\ref{sec:var}).  This weakening apparently affects the dipole component more than the octupole one;  
we however caution that the octupole to dipole intensity ratio, although well constrained in 2010 and 
rather insensitive to minor inversion parameters (like for instance the relative \chisq\ level to which 
the data are fitted or the relative weight assigned to the different reconstructed image quantities in 
the entropy functional), is more uncertain in 2012 with potential values varying from 1.5 to 2.5.  
We thus conclude that the weakening of the overall field is clear, but that the corresponding topological 
change to a more octupolar configuration of the field is only likely.  

We also find that the dipole and octupole components are more or less parallel 
to each other, with respective tilts to the rotation axis of about 25--30\degr\ 
for the dipole (towards phase 0.2 and 0.9 in 2010 and 2012 respectively), and 30 and 0\degr\ for the 
octupole (towards phase 0.5 in 2010).  In particular, the change in the tilt of the octupole component (to 
the rotation axis) between 2010 and 2012 is clear and well constrained by the observations;  the tilts and 
phase of the dipole component is less secure, and likely not accurate to better than 10\degr\ and 0.10 cycle 
respectively.  

The surface brightness distributions that we reconstruct for DN~Tau in 2010 and 2012 both include a dark 
spot close to the pole, but nevertheless feature a number of clear differences.  In 2010, the 
polar spot is essentially monolithic, significantly off-centred from the pole (by about 30\degr\ 
towards phase 0.50) and covering $\simeq$6\% of the overall stellar surface.  
In 2012, the surface brightness distribution of DN~Tau is 
more complex, as readily visible from the RV curve derived from LSD profiles (see top right panel of 
Fig.~\ref{fig:var1}).  It includes not only one cool polar spot almost centred on the pole (with a 
low-contrast appendage towards lower latitudes at phase 0.70), but also a second cool region located 
at intermediate latitudes and centred at phase 0.25;  the overall filling factor associated with this 
brightness distribution is now only $\simeq$4.5\% (including the contribution of both cool features), 
i.e., significantly smaller than that derived from our 2010 observations.  
This clearly demonstrates that the brightness distribution of DN~Tau significantly evolved between 
2010 and 2012;  moreover, since the main cool polar spot overlaps almost perfectly with the main radial 
field magnetic region at both epochs, it suggests that the temporal evolution of the brightness 
distribution mostly reflects the underlying change in the large-scale magnetic topology.  

The overlap of dark spots and magnetic regions also explains a posteriori why the most intense fields 
reconstructed at the surface of DN~Tau are not detected through LSD photospheric profiles 
(see longitudinal field curves at both epochs, bottom panels of Fig.~\ref{fig:var1}).  
Given their low relative brightness, dark spots emit few photons (compared to the surrounding 
unspotted photosphere) and therefore few circularly polarized photons from the magnetic regions 
that they harbor, even for strong magnetic fields;  Stokes $V$ Zeeman signatures from the darkest magnetic 
regions imaged at the surface of DN~Tau are thus too weak to allow retrieving the kG field regions 
they host from LSD profiles alone, even when brightness distributions are reconstructed simultaneously 
with magnetic maps.  Photospheric lines are however crucial for reconstructing the magnetic topology 
at equatorial and intermediate latitudes - a key asset for unravelling the respective contribution 
of the dipolar and octupolar components in the mutipolar field expansion. 

Maps of excess \caii\ emission show a clear accretion region located close to the pole, at latitude 
$\simeq$60\degr\ and phase 0.55 in 2010, and latitude $\simeq$75\degr\ and phase 0.75 in 2012.  
In both cases, this accretion region is located within the darkest spot traced by photospheric LSD 
profiles and covers about 1.5\% of the overall stellar surface.  Zeeman signatures from accretion 
proxies thus ideally complement those from photospheric LSD profiles, allowing us to derive a
well-constrained description of the large-scale field by providing the missing piece of the puzzle 
(the kG fields within the dark polar spot).  The latitudinal shift of the accretion 
region between 2010 and 2012 is readily visible from the rotational modulation of \caii\ and \hei\ 
emission profiles (much smaller in 2012 than in 2010, see Figs.~\ref{fig:var2} and \ref{fig:var3}) 
and further supports our conclusion that the orientation of the overall magnetic topology evolved 
between the two epochs, getting better aligned with the rotation axis in 2012.  
We also note that the accretion region features a low-contrast crescent-shaped appendage in 2010, 
while it is more circular and compact during our second observing run (2012).

\section{Summary \& discussion}
\label{sec:dis}

Our paper presents dual-epoch spectropolarimetric observations of the cTTS DN~Tau aimed at 
unveiling the large-scale magnetic topology present at the surface of the protostar, at looking 
for potential long-term temporal variations of this magnetic topology, and at investigating how 
it impacts accretion from the inner regions of the accretion disc to the stellar surface.  
Being part of the MaPP Large program with ESPaDOnS at CFHT, this analysis comes as a follow-up 
of all similar studies published so far on this subject and brings further information on 
how large-scale fields of cTTSs depend on fundamental parameters such as mass and age.  
More specifically, the current paper focusses on the second lowest mass star of our sample, 
thus complementing the surprising results obtained on the only very-low-mass cTTS 
observed to date \citep[i.e., the 0.35~\msun\ fully-convective protostar V2247~Oph,][]{Donati10} and to get 
further clues on what large-scale fields of cTTSs look like in the very-low-mass regime where rotational 
evolution is different (see Sec.\ref{sec:int}).  

Our spectropolarimetric data of DN~Tau were collected in 2010~Dec and 2012~Dec with ESPaDOnS at CFHT, 
and first allowed us to derive a new estimate of the photospheric temperature (in reasonable agreement 
with previous reports in the literature) and to confirm that DN~Tau is a $\simeq$2~Myr-old fully-convective  
$0.65\pm0.05$~\msun\ star.  Clear circularly-polarized Zeeman signatures are detected at most epochs in 
both LSD profiles of photospheric lines and in the narrow emission features probing ongoing accretion in 
chromospheric layers;  the corresponding longitudinal fields range from --0.3 to 0.2~kG in photospheric 
lines, from --0.1 to 0.7~kG in the emission core of \caii\ lines, and from --0.4 to 1.8~kG in the narrow 
emission profile of \hei\ $D_3$ lines.  Temporal variability of both Zeeman signatures and unpolarized line 
profiles include a significant level of rotational modulation (see Sec.~\ref{sec:var}), with a period fully 
compatible with the most recent literature estimate \citep[i.e., 6.32~d, ][]{Artemenko12};  this confirms 
in particular that the rotation axis of DN~Tau is inclined at $\simeq$35\degr\ to the line of sight.  

Worth noting is that longitudinal fields of DN~Tau as derived from LSD profiles and accretion proxies are not 
systematically of opposite signs \citep[as was the case for, e.g., TW~Hya,][]{Donati11b}, and are even most of 
the time of the same sign in 2012~Dec \citep[as was the case for AA~Tau,][]{Donati10b};  moreover, 
longitudinal field strengths in LSD profiles are, once averaged over the rotation cycle, much lower than those in 
\caii\ lines, by typically a factor of 5, rather than being comparable (as for TW~Hya).  
This qualitatively suggests that the underlying large-scale magnetic topology of DN~Tau is significantly simpler 
than that of TW~Hya - a conclusion confirmed by the subsequent detailed profile modeling.  We also report in this 
paper clear evidence that the large-scale field of DN~Tau is evolving with time between our two observing campaigns, 
both in intensity and orientation, making DN~Tau similar to V2129~Oph and GQ~Lup in this respect \citep{Donati11, Donati12}.  

Thanks to our dedicated tomographic imaging code (tested and optimized for the specific needs of our MaPP observations), 
we derived, from simultaneous fits to our sets of unpolarized and circularly-polarized LSD photospheric profiles and 
accretion proxies, maps of the large-scale field of DN~Tau at both observing epochs, along with surface distributions of 
its photospheric dark spots and its accretion regions.  We find that DN~Tau hosts a mostly poloidal large-scale field, 
largely axisymmetric with respect to the rotation axis and with a polar field strength reaching 1.8 and 1.3~kG at both 
epochs respectively.  The octupolar component of the large-scale field (of polar strength 780 and 600~G in 2010 and 2012 
respectively) is larger than the dipole component (530 and 300~G in 2010 and 2012) by a factor of 1.5 (in 2010) to 
2.0 (in 2012);  it implies that the overall topology of DN~Tau remains rather simple, 
featuring in particular no high-contrast polarity reversals over the visible hemisphere.  
This makes DN~Tau the cTTS harbouring the simplest magnetic topology of all MaPP stars observed to date, 
after that of AA~Tau (whose large-scale field is dominated by a dipole at least $\simeq$4 times stronger than the 
octupole component).  The orientation of the octupole component (roughly parallel to the dipole component) is 
clearly changing with time;  tilted at $\simeq$30\degr\ in 2010, it is almost aligned with the rotation axis in 2012, 
further demonstrating the genuine variability of the large-scale fields of cTTSs on timescales of years.  

\begin{figure*}
\includegraphics[scale=0.65,angle=-90]{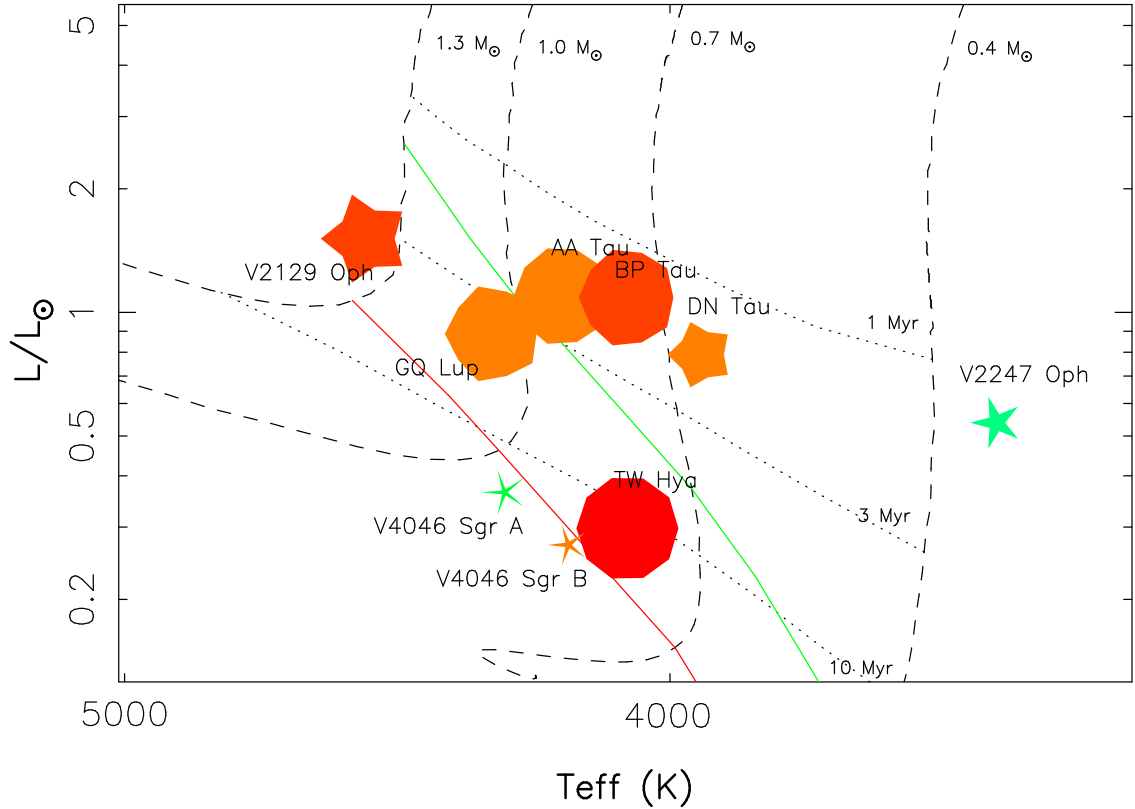}
\caption[]{Basic properties of the large-scale magnetic topologies of cTTSs, as a function of their locations in the HR diagram.  
Symbol size indicates relative magnetic intensities, symbol color illustrates field configurations (red to blue for purely 
poloidal to purely toroidal fields), and symbol shape depicts the degree of axisymmetry of the poloidal field component 
(decagon and stars for purely axisymmetric and purely nonaxisymmetric poloidal fields, respectively).  The large-scale 
field of DN~Tau is weaker than that of AA~Tau and BP~Tau, but still mainly poloidal and largely axisymmetric (the non-decagonal 
shape of the symbol being due to the 30\degr\ tilt of the octupole field in 2010).  
The PMS evolutionary tracks and corresponding isochrones \citep{Siess00} assume solar metallicity and include convective 
overshooting.  The full lines depict where models predict cTTSs start developing their radiative core (green line) and when 
their convective envelope is thinner than 0.5~\rstar (red line), as they contract towards the main sequence. 
This figure is best viewed in color.  }
\label{fig:hrd2}
\end{figure*}

The large-scale field we recover for DN~Tau features both similarities and differences with those of the other cTTSs 
observed in the MaPP framework (see Fig.~\ref{fig:hrd2}).  It is largely poloidal, with an octupole to dipole intensity 
ratio smaller than 2, like other fully-convective cTTSs located nearby in the HR diagram 
(namely AA~Tau and BP~Tau);  it is however quite different from that of the very-low-mass cTTS V2247~Oph 
(the closest fully-convective neighbour of DN~Tau on the very-low-mass side of the MaPP sample) whose large-scale field 
includes a significant toroidal component and a mostly non-axisymmetric poloidal component.  
However, the large-scale field of DN~Tau is significantly weaker than that of its higher mass neighbours AA~Tau and BP~Tau 
and more comparable (in strength) to that of V2247~Oph despite the difference in topology.  We think that these discrepancies are 
unlikely attributable to temporal variations of the large-scale fields resulting from non-stationary dynamos - 
usually much smaller in amplitude than this interpretation would require;  
we rather speculate that these differences illustrate the expected progressive topological transition between low-mass 
and very-low-mass fully-convective cTTSs, similar to that observed between mid-M and late-M main-sequence dwarfs 
\citep[e.g.,][]{Morin08b, Morin10}.  In both cases, our observations bring yet further support on the fact that 
large-scale fields of cTTSs are produced through non-stationary dynamo processes.  

\cite{Gregory12} reasoned that, as with late-M-dwarfs, the lowest mass cTTSs may host a variety of field topologies, 
some being significantly more complex than what is found for the more massive fully-convective stars such as AA~Tau \citep{Donati10b}.  
They argued that a bistable dynamo regime would exist somewhere between 0.2~\msun\ as a lower limit \citep[the mass below which 
bistable dynamo behaviour has been observed for main sequence M-dwarfs,][]{Morin10} and 60\% of the fully-convective limit at 
any given PMS age as an upper limit (as fully-convective stars are found below 0.35~\msun\ on the main sequence, and 0.2~\msun\ is 
$\simeq$60\% of this mass).  As we go to successively older ages, the mass below which fully-convective cTTSs are found decreases 
\citep[see Fig.~\ref{fig:hrd2} and equation B1 in][]{Gregory12}.  At the age of DN~Tau, $\simeq$1.7~Myr, the fully convective limit is $\simeq$1.2~\msun.  
With a mass of 0.65~\msun, DN~Tau therefore lies below the upper limit where \citet{Gregory12} argued that bistable dynamo behaviour 
may be found.  This may explain the weaker field of DN~Tau relative to the more massive fully convective stars AA~Tau and BP~Tau. 

We report as well the presence of a dark photospheric region overlapping the main magnetic pole of DN~Tau at both epochs, 
as in all other fully- or mainly-convective cTTSs with masses larger than 0.6~\msun\ observed so far with MaPP.  
Similarly, DN~Tau also features an accretion region at chromospheric level, located close to the magnetic pole 
and within the dark photospheric spot, strongly suggesting that accretion from the inner disc regions is occurring 
mostly poleward in DN~Tau as well.  We note that the accretion region features a crescent shape in 2010, i.e., 
when the octupole to dipole intensity ratio is smallest, while it shows a more circular shape in 2012, i.e., 
when the octupole to dipole intensity ratio is largest (than in 2010);  this behaviour is in qualitative agreement with 
what we expect for the shape of accretion spots from theoretical simulations of magnetospheric accretion 
\citep[e.g.,][]{Romanova04b, Romanova11}.  

Given the logarithmic mass accretion rate (of $-9.1\pm0.3$ in \mspy) that we infer from emission fluxes of conventional 
accretion proxies, we obtain that the large-scale field of DN~Tau should be able to disrupt the accretion disc up to 
a radius of $\rmag\simeq5.9$~\rstar\ (0.052~au) in 2010 and $\simeq4.3$~\rstar\ (0.038~au) in 2012 
\citep[assuming an average dipole strength of 0.53 and 0.30~kG in 2010 and 2012 respectively, and using the analytical 
formula of][]{Bessolaz08}\footnote{As the dipole component drops most slowly with distance from the star, it alone provides 
an adequate approximation of \rmag\ \citep{Adams12}.  The field strength along the magnetic loops truncating the disc 
does, however, depart from that expected from a pure dipole loop.  In particular, the field strength at the base of the 
magnetic loop (i.e., that probed by accretion proxies such as \hei\ D$_3$) is significantly larger than that corresponding 
to the dipole component alone. }.  When \rmag\ is compared to the corotation radius $\rcor\simeq6.6$~\rstar\ (or 0.058~au), 
at which the Keplerian period equals the stellar rotation period, we find that $\rmag/\rcor$ is equal to 
$\simeq0.90$ and $\simeq0.65$ in 2010 and 2012 respectively.  We caution that this estimate of $\rmag/\rcor$ 
\citep[assuming a sonic Mach number at disc mid-plane of $m_{\rm s}\simeq$1, see][]{Bessolaz08} is likely an upper 
limit only, numerical simulations suggesting a potential overestimate of $\simeq$20\% \citep{Bessolaz08}.  

Whereas spin-down effects caused by star/disc magnetic coupling can be effective and partly counteract 
the accretion torque even when $\rmag/\rcor\simeq$~0.5--0.8, recent numerical simulations suggest that they can 
only overcome both the accretion and stellar contraction torques for values of $\rmag/\rcor\simeq$~0.8--1, 
at which outflows are generated in the form of magnetospheric ejections through a propeller-like mechanism 
\citep[e.g.,][]{Romanova04, Zanni13}.  Accretion-powered stellar winds \citep{Matt05, Matt08} are also 
unlikely to succeed at spinning down DN~Tau;  the wind spin-down torque would indeed require the mass ejection 
rate of DN Tau to be several times larger than the mass accretion rate we determined - obviously not compatible with the 
fact that DN Tau is not in a propeller-like regime (where most of disc material is ejected and only a small fraction gets 
accreted).  This suggests that DN~Tau already entered a phase of spin-up, and potentially explains why it is rotating 
slightly faster (with a period of 6.32~d) than prototypical slowly-rotating cTTSs like AA~Tau.  

Conversely to the case of non-fully-convective cTTSs like TW~Hya or V2129~Oph, the reason of the spin-up of DN~Tau 
(and of the correspondingly weaker dipole component) is not attributable to a recent change in the internal stellar 
structure.  We rather speculate that this is caused by a different regime of dynamo processes in very-low-mass stars, 
having much harder times at producing strong, aligned dipolar fields, with DN~Tau and V2247~Oph providing examples of a 
smooth progressive transition to this new dynamo regime;  this could qualitatively explain at the same time the specific 
rotational evolution that very-low-mass protostars are subject to.  
More spectropolarimetric observations like those reported here, especially in the low-mass, fully-convective 
region of the HR diagram, are required to confirm our speculations.  In particular, SPIRou, the nIR spectropolarimeter~/ 
high-precision velocimeter presently designed as a next generation instrument for CFHT, should be particularly efficient 
at exploring this region of the HR diagram to investigate in more details bistable dynamos of cTTSs.

\section*{Acknowledgements}
We thank an anonymous referee for valuable comments and suggestions that improved the manuscript.
This paper is based on observations obtained at the Canada-France-Hawaii Telescope (CFHT), operated by the National
Research Council of Canada, the Institut National des Sciences de l'Univers of the Centre
National de la Recherche Scientifique (INSU/CNRS) of France and the University of Hawaii, 
and at the T\'elescope Bernard Lyot (TBL), operated by INSU/CNRS.
The ``Magnetic Protostars and Planets'' (MaPP) project is supported by the
funding agencies of CFHT and TBL (through the allocation of telescope time)
and by INSU/CNRS in particular, as well as by the French ``Agence Nationale
pour la Recherche'' (ANR).
SGG acknowledges support from the Science \& Technology Facilities Council (STFC) via an Ernest Rutherford Fellowship [ST/J003255/1]. 
SHPA acknowledges support from CNPq, CAPES and Fapemig.
FM acknowledges support from the Millennium Science Initiative (Chilean Ministry of Economy), through grant ``Nucleus P10-022-F''.

\bibliography{dntau}
\bibliographystyle{mn2e}
\end{document}